\documentclass[pra,aps,showpacs,preprint]{revtex4}
\usepackage{dcolumn}
\input epsf
\usepackage{epsfig}
\begin{document}

\title{Measuring Atmospheric Neutrino Oscillations with Neutrino
Telescopes}

\vspace*{-24pt}

\author{Ivone F.\ M. Albuquerque }\thanks{Electronic mail: 
                IFAlbuquerque@lbl.gov}
\affiliation{Department of Astronomy \& Space Sciences Laboratory,
University of California, Berkeley, \ CA 94720. }

\author{George F. Smoot}\thanks{Electronic mail: 
                {GFSmoot@lbl.gov}}
\affiliation{Department of Physics, Lawrence Berkeley National Laboratory \\
\& Space Sciences Laboratory.
University of California, Berkeley,\ CA 94720.}

\date{28 March 2001}

\begin{abstract}
Neutrino telescopes with large detection volumes can demonstrate 
whether the current indications of neutrino oscillation are correct or 
if a better description can be achieved with non-standard alternatives.
Observations of contained muons produced by atmospheric neutrinos
can better constrain the allowed region for oscillations or determine
the relevant parameters of non-standard models.
We analyze the possibility of neutrino telescopes measuring atmospheric 
neutrino oscillations. We suggest adjustments to improve this potential.
An addition of four densely-instrumented strings to the AMANDA II detector
makes oscillation observations feasible.
Such a configuration is competitive with current and proposed 
experiments.
\end{abstract}


\maketitle

\section{Introduction}

Experimental observations (Super-Kamiokande \cite{superk}, 
Kamiokande \cite{kamioka}, IMB \cite{imb} and Soudan \cite{soudan})
for atmospheric muon and electron neutrinos have found that the
ratio of the number of these neutrino species does not agree with theoretical
prediction. 
All experiments measuring the flux of solar neutrinos observe a deficit 
compared with the solar model predictions \cite{Bahcall}. 
The ratio of muon to electron events observed in atmospheric neutrino 
interactions is measured by most experiments to be less than expected 
from models of cosmic ray interactions in the atmosphere \cite{Fogli}.
Neutrino oscillation can explain these results. 
The measurement of the up/down asymmetry of this ratio by the Super-Kamionkande 
collaboration \cite{superk} is generally considered to be the strongest evidence
for neutrino oscillations.
As oscillations would most likely imply that neutrinos have mass, 
many researchers have fitted the available data for the mass difference among 
different neutrino species. 

While the cumulative evidence for neutrino oscillations is very striking,
a definitive proof that the observed anomalies are actually due to neutrino 
oscillations is still missing. 
In particular the current observations of atmospheric neutrinos are consistent
with the hypothesis of maximal $\nu_\mu$ oscillations, 
but do not exclude some alternative unconventional explanations, 
such as neutrino decay \cite{Barger}, micron-scale extra dimensions \cite{Barbieri} or 
quantum decoherence in propagation \cite{Lisi}.

Cosmic rays traveling to Earth will interact with its atmospheric nuclei.
These interactions produce hadrons which decay eventually into atmospheric neutrinos.
Up to neutrino energies of about 100 GeV, the main result from these 
interactions is pion production $\pi^{+} (\pi^-)$. These will 
decay into $\mu^{+} + \nu_{\mu}$ ($\mu^{-} + \overline{\nu}_{\mu}$)
almost 100\% of the time. The
secondary $\mu^+ (\mu^-)$ will decay into $e^{+} + \nu_e + \overline{\nu}_{\mu}$
($e^{-} + \overline{\nu}_e + \nu_{\mu}$) which will give a 
$\nu_{\mu}+\overline{\nu}_{\mu}$
to $\nu_e+\overline{\nu}_e$ ratio (r) of 2:1. 

However, depending on the cosmic ray energy and
where it interacts in the Earth atmosphere, 
the muon might not decay before reaching the ground and corrections 
to this ratio have to be included. Another effect to be included is the Earth's magnetic field action on lower energy particles. 
These effects have been modeled and simulated and one looks for a 
ratio ($R$) that relates $r_{exp}$ 
from experimental data to $r_{sim}$ 
from simulations ($R = r_{exp}/r_{sim}$). 
In this way one can reduce the bias introduced by uncertainties 
in understanding these effects. 

Super-Kamiokande, Kamiokande, IMB and Soudan experiments \cite{superk,kamioka,imb,soudan}
found the ratio R to be lower than expected, with fewer $\nu_{\mu}$'s. 
Frejus and Nusex experiments \cite{frejus1,nusex} have not found any 
anomaly in R.

If neutrinos have mass, one type of neutrino can oscillate into another. 
The probability (P) for atmospheric neutrino oscillations, 
in a two neutrino mixing scenario, is given by \cite{kayser}:

\begin{equation}
P = \sin^2{2\theta}\sin^2\left(1.27 \frac{L}{E} \Delta m^2\right),
\label{eq:oscprob}
\end{equation}
where $\theta$ is the mixing angle, $\Delta m^2$ is the difference of the
squared mass  for two types of neutrinos
in $eV^2$, $L$ is the distance traveled by the original neutrino in km and 
$E$ is the neutrino energy in GeV. 

The Super-Kamiokande analysis \cite{superk} constrains $\Delta m^2$ 
from $5\times 10^{-4}$ to $6 \times 10^{-3} eV^2$ and $\theta$ 
to greater than 0.82 at 90\% confidence level. 
The most probable solution is $\Delta m^2 = 3.5 \times 10^{-3} eV^2$ and 
$\sin^2{2\theta} = 1$ (full mixing). 
Also the oscillation of a muon neutrino into a tau neutrino is favored 
over an oscillation into a sterile neutrino \cite{sktaufavor}. 

If the atmospheric neutrino result is joined with the solar and 
short baseline beam measurements, 
they indicate evidence of a non-zero $\Delta m^2$. 
However, new neutrino oscillation experiments are needed to precisely 
measure $\Delta m^2$ as well as to decide 
among experimental results that are in disagreement. 
As an example, 
the Super-Kamiokande \cite{superk} results barely overlap with Kamiokande 
\cite{kamioka} results.

As the detector area and volume covered by current or proposed 
high-energy astrophysical neutrino telescopes are large  
compared to underground detectors, it is possible for these experiments 
to contribute to the understanding of atmospheric neutrino oscillations. 
In this paper we investigate the possibility of AMANDA \cite{amanda} and 
IceCube \cite{ice3} detectors measuring atmospheric neutrino oscillations.
Similar analyses would apply 
for the ANTARES \cite{antares} and NESTOR \cite{nestor} detectors. 

\section{Measurement of Atmospheric Oscillations}

\begin{figure}
\centering\leavevmode \epsfxsize=300pt \epsfbox{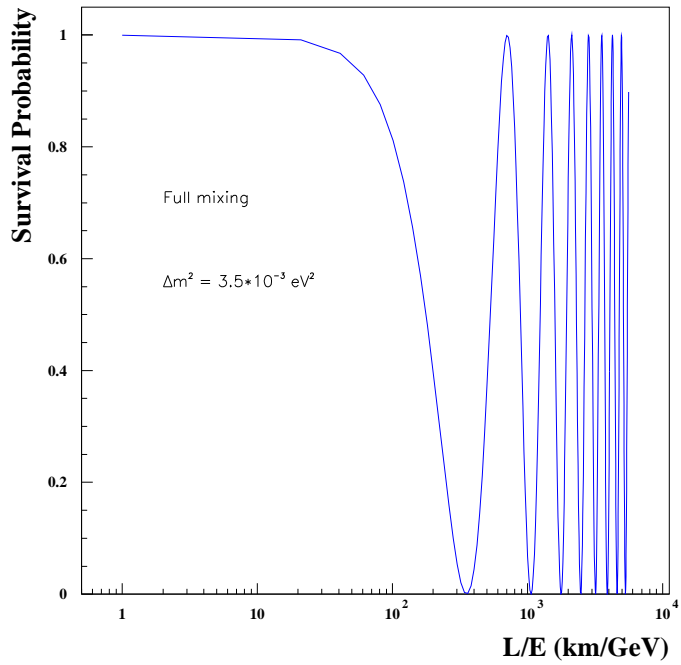}
\caption{Neutrino survival probability versus $L/E$. Full neutrino mixing 
($\sin^2 2 \theta = 1$) and 
\protect$\Delta m^2 = 3.5 \times 10^{-3}\; eV^2$ is assumed.}
\label{fig:osc}
\end{figure}

Figure~\ref{fig:osc} shows the neutrino flavor survival probability ($P_S$) versus 
the ratio of the distance traveled by the neutrino and its energy ($L/E$). 
This probability is given by $1-P_\sim$ where $P_{\sim}$
is the neutrino oscillation probability given in Equation~\ref{eq:oscprob}. 

As one can experimentally determine the number of events with 
energy E as a function of the zenith angle ($\theta_Z$), 
a plot of the survival probability versus $\cos\theta_Z$ 
provides a more convenient indication of what could be accomplished by 
neutrino telescopes.

Equation~\ref{eq:oscprob} shows that the larger the distance the neutrino
travels, the greater the possibility of oscillation. 
For energies of 10 GeV and above only upward going neutrinos,
those which come from below the detector after crossing the Earth, 
will have a significant probability of oscillating. 

We define $\theta_Z$ as the angle from the detector axis 
(defined by a line from the detector to the center of the Earth) 
to the direction of the particle arrival at the detector 
\footnote{\protect$\theta_Z = 0$ degree corresponds 
to upward going neutrinos and \protect$\theta_Z = 180$ degrees 
corresponds to downward going neutrinos.}. 
For a detector located at a distance $R_d$ from
the center of the Earth, 
the distance traveled (L) by the neutrino will be:

\begin{equation}
L = R_d \cos\theta_Z + \sqrt{R_\nu^2 - R_d^2 + R_d^2 \cos^2\theta_Z}
\label{eq:len}
\end{equation}
where $R_\nu$ is the radius from the Earth's center at which the neutrino is produced, $R_\nu  \simeq  R_\oplus + 15$~km, 
where $R_\oplus$ is the Earth radius. 
This equation reduces to $L = R_d + R_\nu \sim 2 R_\oplus$
for $\cos\theta_Z = 1 $ (vertical upcoming neutrinos), 
to $L = \sqrt{R_\nu^2 - R_d^2}$ for $\cos^2\theta_Z = 0 $ (horizontal neutrinos), 
and $L = R_\nu - R_d \sim 15$~km 
for $cos\theta_Z = -1$ (vertical downgoing neutrinos).

At the energies we are considering (10-100 GeV) only upwards going neutrinos will 
have a significant probability of oscillating. 
Thus the neutrino production point in the atmosphere 
will not be significant when compared to the length traveled
through the Earth.

\begin{figure}
\centering\leavevmode \epsfxsize=300pt \epsfbox{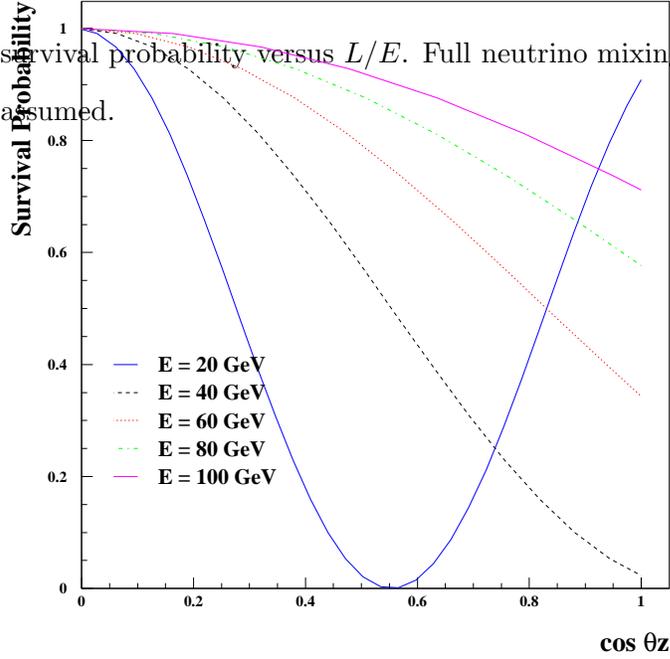}
\caption{Neutrino survival probability versus cosine of the zenith
angle for different neutrino energies as labeled. 
Full neutrino mixing and \protect$\Delta m^2 = 3.5 \times 10^{-3}\; eV^2$ is 
assumed. 
The probability is given for an idealized detector 2 km deep in the Earth.
It indicates that a neutrino telescope needs a low energy threshold 
(less than 30 GeV) to be able to measure atmospheric oscillations well.}
\label{fig:surv}
\end{figure}

Figure~\ref{fig:surv} shows $\cos\theta_Z$ versus the survival probability 
as a function of energy.
We assume that the detector is 2 km deep in the Earth (ice or water).
It can be seen that it is easier to detect oscillations using lower energy
neutrinos. Therefore to measure neutrino oscillations the telescope energy 
threshold cannot be too high and the energy resolution must be good. 
However, for high-energy neutrino telescopes the sensing elements are 
placed far apart to gain detector volume and the trigger energy threshold 
is usually set high to avoid atmospheric muon and neutrino background. 
The detector design has to be optimized, if it is to be used for both of 
these two types of observations.

\begin{figure}[t]
\centering\leavevmode \epsfxsize=300pt \epsfbox{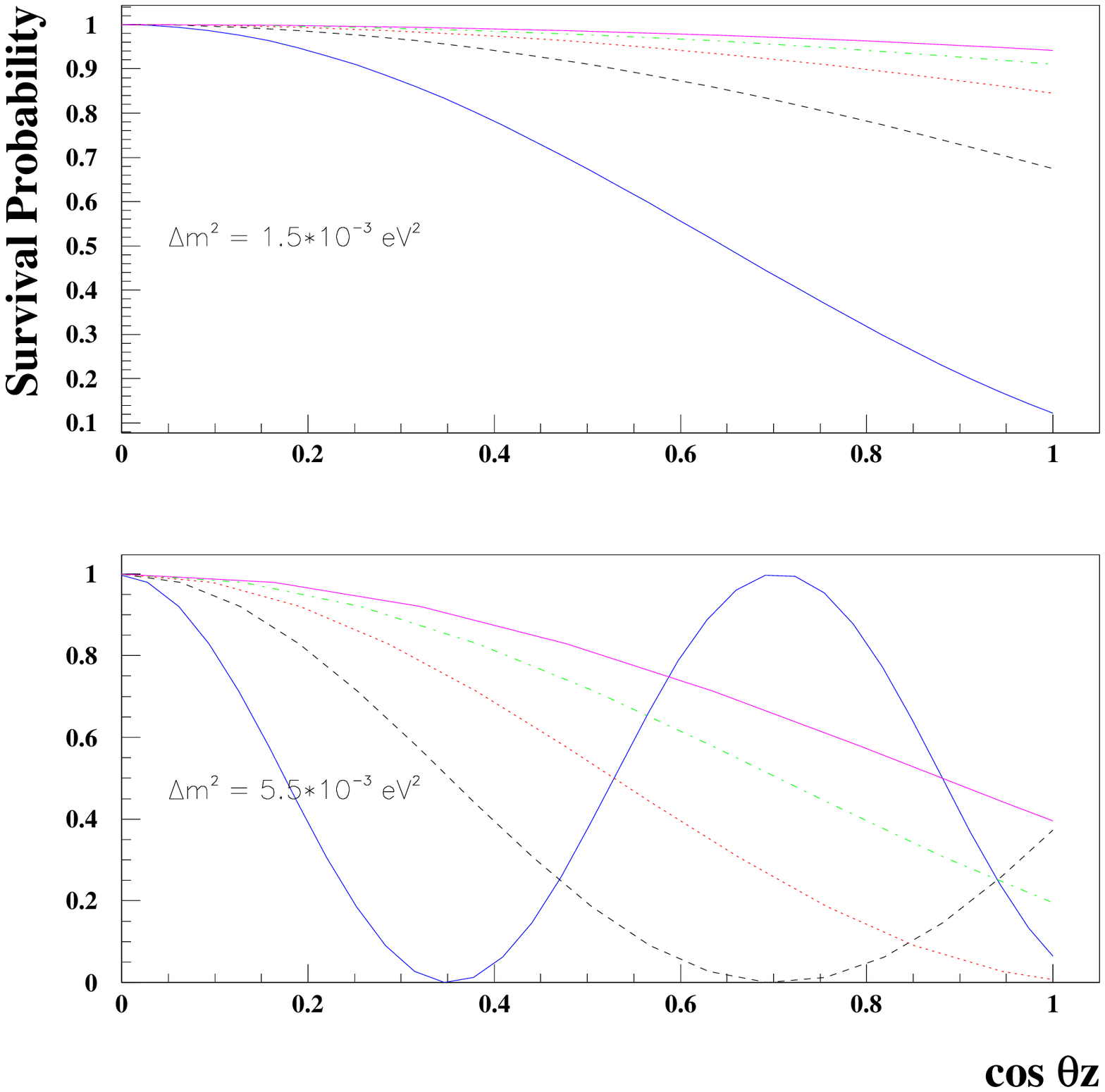}
\caption{Neutrino survival probability versus cosine of the zenith
angle for different neutrino energies (as labeled in figure \protect\ref{fig:surv}).
The probability is given for a detector 2 km deep in the Earth.
Full mixing is assumed and $\Delta m^2 = 1.5 \times 10^{-3}$,
$\Delta m^2 = 5.5 \times 10^{-3}\; eV^{2}$ as labeled. One can see that
even for larger $\Delta m^2$ the neutrino telescope energy threshold has to be equal or
less than 30 GeV.}
\label{fig:oscdm2}
\end{figure}

Both Figure~\ref{fig:osc}~and~\ref{fig:surv} assume full mixing and 
$\Delta m^2 = 3.5 \times 10^{-3}\; eV^2$. 
In Figure~\ref{fig:oscdm2} we show the same dependencies as 
in Figure~\ref{fig:surv}, but for different values of $\Delta m^2$.
From this figure one can see that, depending on the energy threshold,
it may be possible to better constrain $\Delta m^2$ with a neutrino telescope. 
To understand whether or not this measurement can be done, 
one has to take into account the atmospheric neutrino flux, 
the rate of contained events and the energy resolution of the detector.

\begin{figure}
\centering\leavevmode 
\epsfig{figure=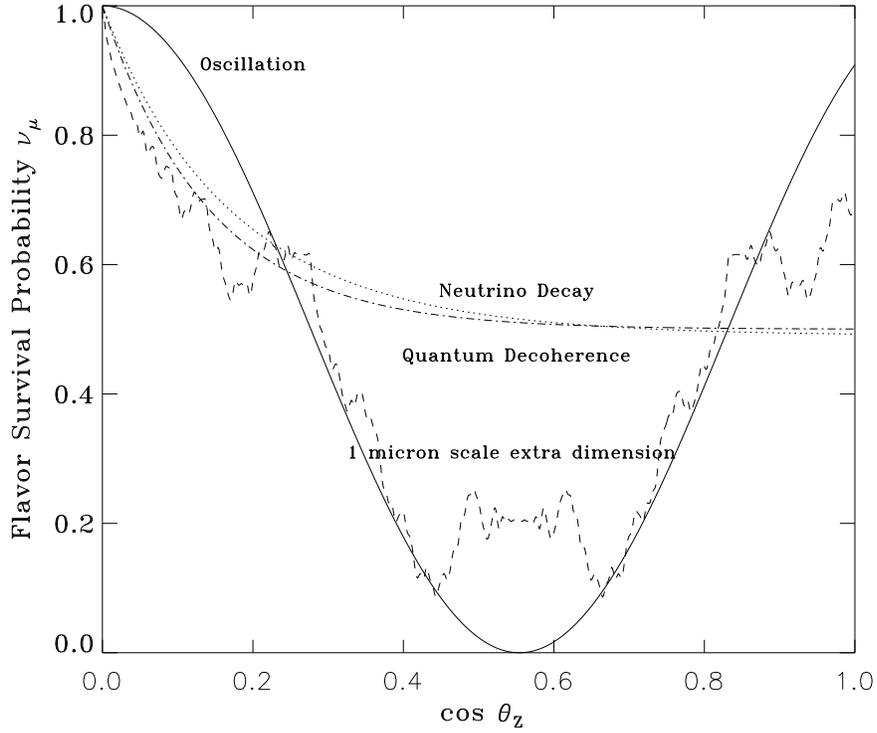,width=300pt,angle=90}
\caption{Four theoretical scenarios that are consistent with the Super-Kamionkande 
results. 
(1) Solid line shows best fitted standard neutrino oscillations. 
Neutrino survival probability versus $\cos \theta_Z$ for 20 GeV
neutrinos crossing through the Earth ($L = $ Earth Diameter). 
Full neutrino mixing and \protect$\Delta m^2 = 3.5 \times 10^{-3}\; eV^2$ is 
assumed. 
(2) Dashed line is effect of additional dimension with characteristic radius of about
a micron \cite{Barbieri}.
(3) Dotted line is effect of a decaying neutrino \cite{Barger}.
(4) Dashed dotted line is the effect of quantum decoherence \cite{Lisi}.}
\label{fig:neuoscd}
\end{figure}

Figure~\ref{fig:neuoscd} shows the neutrino oscillation pattern for 20 GeV
neutrinos superimposed by other possible explanations which are compatible with
the Superkamiokande \cite{superk} results. Measuring this oscillation pattern can
allow to distinguish among different scenarios for the muon neutrino deficit. 

\section{Atmospheric Neutrino Flux}
\label{sec:nuflux}

To determine the number of events expected to be detected, 
one has to convolve the survival probability with the atmospheric neutrino 
flux.
We determine this flux versus $\cos{\theta_Z}$ from the flux calculated by Volkova
\cite{volkova} and compare it with other calculations \cite{honda,agrawal}.

Volkova derives the atmospheric neutrino flux from the decay of light mesons
($K, \pi$) and $\mu$'s and from the decay of short-lived particles (prompt decay)
which mainly consist of charm particles. The latter will only be significant at
higher energies (around a PeV) but we include it for completeness.

The differential energy spectra of atmospheric muon neutrinos from light mesons and
muons can be approximated by \footnote{Although Volkova \cite{volkova} uses a lower
limit of \protect$10^2$~GeV for this approximation, the extrapolation down to
10 GeV is in very good agreement with the flux for lower energies which can be
found in table~II of her paper.}:

\begin{eqnarray}
\left(\frac{dN}{dE}\right)_{\rm light} & = & 2.85 \times 10^{-2} E^{-2.69} \left( \frac{1}{1+6E/E_\pi(\theta_Z)}
+ \frac{0.213}{1+1.44E/E_k(\theta_Z)}\right),\nonumber \\
& & 10 \leq E < 5.4 \times 10^5\; {\rm GeV} \\
 & = & 0.48 E^{-4.04} (E_\pi(\theta_Z) + 0.89 E_K(\theta_Z)), \hspace*{.5cm}
E\geq 5.4 \times 10^5\; {\rm GeV}, \nonumber
\label{eq:nufll}
\end{eqnarray}
where $E_\pi(\theta_Z)$ and $E_K(\theta_Z)$ are the $\pi$ and $K$ critical
energies, $E$ is the neutrino energy and the spectra is given in units of
neutrinos per $\rm{cm^2 \; sec\; sr\; GeV}$. The critical energy is the one 
for which the
probability for a nuclear interaction in one nuclear mean free path
equals the decay probability in the same path. It depends on the zenith angle since
the atmosphere density varies with depth and horizontal
events cross more dense regions than the vertical ones. 

The spectra for prompt neutrinos can be approximated by \cite{volkova}:

\begin{eqnarray}
\left(\frac{dN}{dE}\right)_{\rm prompt} & = & 2.4 \times 10^{-5} E^{-2.65}, \hspace*{.3cm}
E < 2.3 \times 10^6\; {\rm GeV}\\
 & = &  3.9 \times 10^{-3} E^{-3}\hspace*{.3cm}, \hspace*{.3cm}
E \geq 2.3 \times 10^6\; {\rm GeV}. \nonumber
\label{eq:nuflp}
\end{eqnarray}
 
\begin{figure}
\centering\leavevmode \epsfxsize=300pt \epsfbox{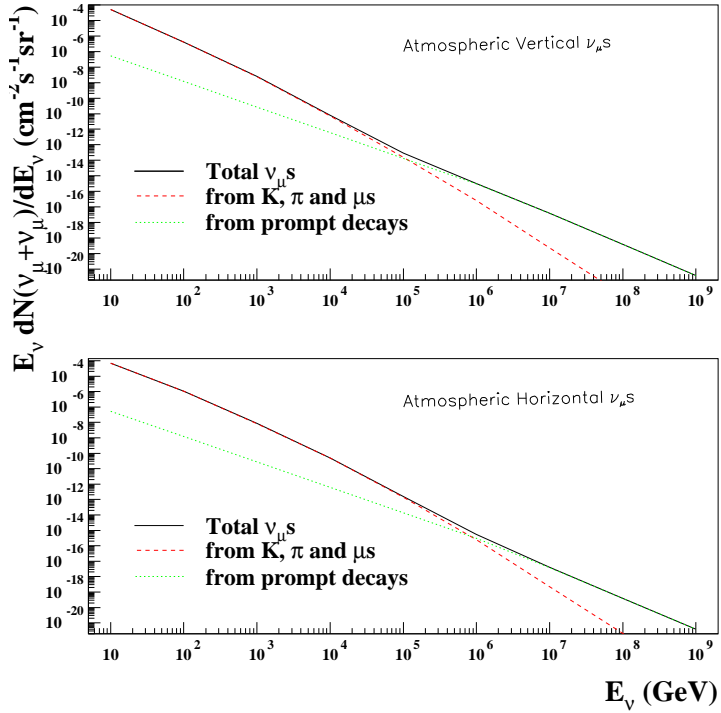}
\caption{Atmospheric vertical and horizontal (as labeled) muon neutrino energy spectra
based on \protect\cite{volkova}. Also shown the contribution from $K,\,\pi$ and $\mu$'s and 
from prompt decays.}
\label{fig:nuflux}
\end{figure}

\begin{figure}
\centering\leavevmode \epsfxsize=300pt \epsfbox{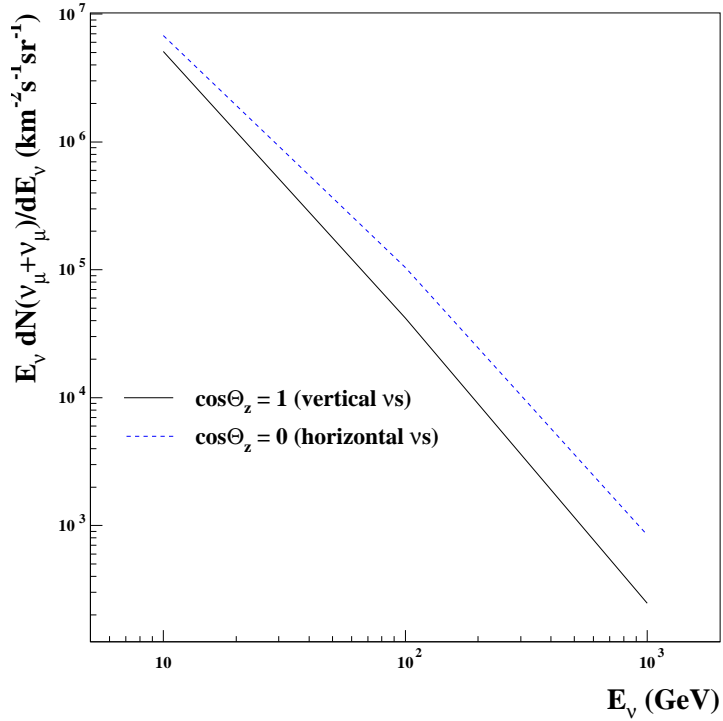}
\caption{Atmospheric vertical and horizontal (as labeled) muon neutrino energy spectra.
Note that it is plotted per ${\rm km^2}$ instead of ${\rm cm^2}$ as in the previous 
figure.
The flux of horizontal neutrinos is higher than vertically down going neutrinos
since the muons have more opportunity to decay.}
\label{fig:nufluxcut}
\end{figure}

In Figure~\ref{fig:nuflux} we show the vertical and horizontal neutrino energy 
spectra and the contributions from $K,\,\pi$ and $\mu$'s and from prompt decays.
One can see that the prompt contribution is negligible at lower energies.
In Figure~\ref{fig:nufluxcut} we expand the lower energy region and plot
the total flux for vertical ($\cos\theta_Z = 1$) and horizontal ($\cos\theta_Z = 0$)
spectra. As expected the number of horizontal neutrinos is slightly higher.


We compare this flux with that obtained by \cite{honda} and 
\cite{agrawal}.
The largest discrepancy between these calculations is of about 15\% 
(see Figure~14 of ref.~\cite{honda} and Figure~7 of ref.~\cite{agrawal}). 
In the energy range of interest to our work, the Volkova
spectrum is an underestimate compared to these other spectra. 
We will use the atmospheric neutrino flux based on the Volkova spectra.
This is a more conservative spectrum and having a larger flux will only enhance the
possibility of measuring neutrino oscillations.
Also important is to compare the slope of these spectra and Figure~7 of 
ref.~\cite{agrawal}) shows that for energies between 10 and 100 GeV the
difference between the slope of the Volkova spectrum and of the spectrum
determined in \cite{agrawal} or \cite{honda} is at maximum 1\%.

\begin{figure}
\centering\leavevmode \epsfxsize=300pt \epsfbox{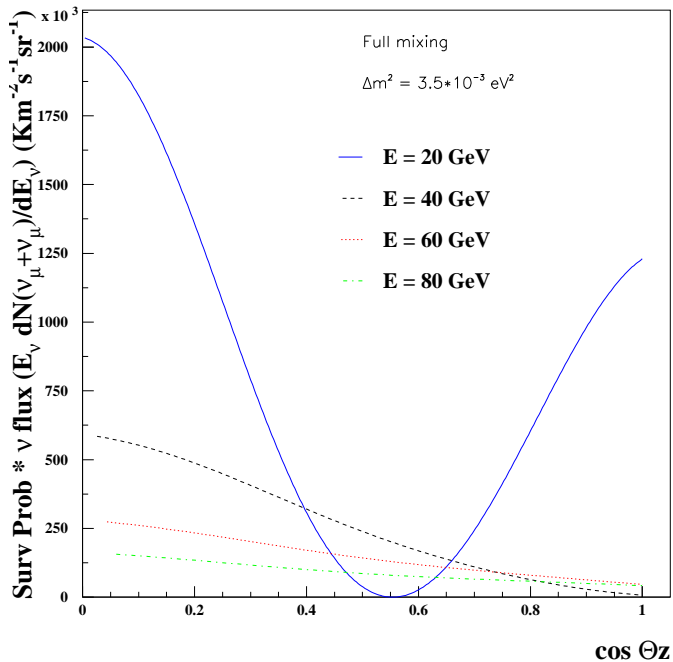}
\caption{Muon neutrino survival probability weighted by atmospheric muon neutrino energy 
spectrum. 
As expected, neutrino oscillations are enhanced at lower energies by a 
combination of higher flux and more rapid oscillation probability.}
\label{fig:survflux}
\end{figure}

We now proceed to convolve the survival probability for upward going neutrinos shown in 
Figure~\ref{fig:surv}
by the flux obtained from the above analysis. 
This is shown in Figure~\ref{fig:survflux}.

As expected, neutrino oscillations are enhanced at lower energies.

\section{Contained Events}

When muon neutrinos undergo a charged current interaction, they produce a muon. 
This will propagate losing energy and eventually comes to rest and decays.
If the entire event happens inside the detector (a ``contained'' event) it is 
much easier to reconstruct the event than if it is not fully contained.

The probability that a neutrino suffers a charge current interaction is given by:

\begin{equation}
P_{\rm conv} = n \sigma_{\rm cc} l
\label{eq:pconv}
\end{equation}
where $n$ is the number density of nucleons of the medium transversed by the neutrino, 
$\sigma_{\rm cc}$ is the charged current cross section for a neutrino nucleon interaction
and $l$ is the distance traveled by the neutrino inside the detector.
We determine $\sigma_{\rm cc}$ according the CTEQ4-DIS distributions, as described in \cite{quigg}. For energies below 10 GeV a correction on the cross
section due to quasi-elastic and resonant effects \cite{lip} should be
included. As our analysis starts with energies above 10 GeV we do not
include these corrections.

The number of contained events ($N_{\rm cont}$) is therefore given by:

\begin{equation}
N_{\rm cont} = P_{\rm conv} \varphi_\oplus A = n \sigma_{\rm cc} \varphi_\oplus V, 
\label{eq:ncont}
\end{equation}
where $\varphi_\oplus$ is the flux of upward going neutrinos, $A$ is the 
detector area and $V$ is the detector volume for contained neutrino 
interactions. 

The neutrino flux determined in the previous section will suffer a minor
attenuation when going through the Earth. 
This attenuation is not significant at these low energies
but we will include it for completeness. 
The differential flux is given by:

\begin{equation}
\frac{d\varphi}{dx} = -n \sigma_{\rm cc} \varphi
\label{eq:dfatt}
\end{equation}
where $\varphi$ is the atmospherical neutrino flux, $x$ is the distance traveled by the neutrino. 
The atmospheric neutrino flux after transversing the Earth will be given by:

\begin{equation}
\varphi_{\oplus} = \varphi_0 e^{-\int n \sigma_{\rm cc} dx} 
\simeq  \varphi_0 e^{- n_\oplus \sigma_{\rm cc} L} 
\sim \varphi_0 e^{-2 {n_\oplus} \sigma_{\rm cc} {R_\oplus}
 \cos{\theta_Z}}
\end{equation}
where $R_\oplus$ is the Earth radius and $\varphi_0$ is the initial neutrino flux, the approximations are for constant density and negligible detector depth, respectively.

The volume of the detector available for the neutrino to interact and produce 
a contained muon  depends on the muon range ($R_\mu$). 
Making the approximation that the detector
has a cylindrical shape, the effective volume V will be given by:

\begin{equation}
V = \frac{1}{2} h D_d^2 \arcsin\left(\sqrt{1 - \frac{R_\mu^2}{D_d^2}\sin^2\theta_Z}\right)
    \left( 1 - \frac{R_\mu}{h}|\cos\theta_Z|\right)
\label{eq:vol}
\end{equation}
where $h$ is the detector height and $D_d$ is the detector diameter.

For muons with energies of tens of GeV, the average muon range can be
determined analytically by the equation \cite{pdg}:

\begin{equation}
R_\mu = \frac{1}{b} \ln \left( 1 + \frac{b}{a} E_0 \right)
\label{eq:muran}
\end{equation}
where $a$ and $b$ are respectively ionization and radiation loss parameters and $E_0$ is the initial muon energy. 
The average $E_0$ is equal to the neutrino energy minus
the average energy loss in a charged current interaction. 
The fractional energy loss is given in \cite{quigg96} and 
for neutrinos energies between 10-100 GeV is 0.48. 
At these energies $a$ and $b$ can be approximated to constant values 
where $a = 2 \times 10^{-3} {\rm GeV\,cm^2\,/\,g}$ and 
$b = 4 \times 10^{-6} {\rm cm^2\,/g}$ \cite{pdg}.

\begin{figure}
\leavevmode \epsfxsize=225pt \epsfbox{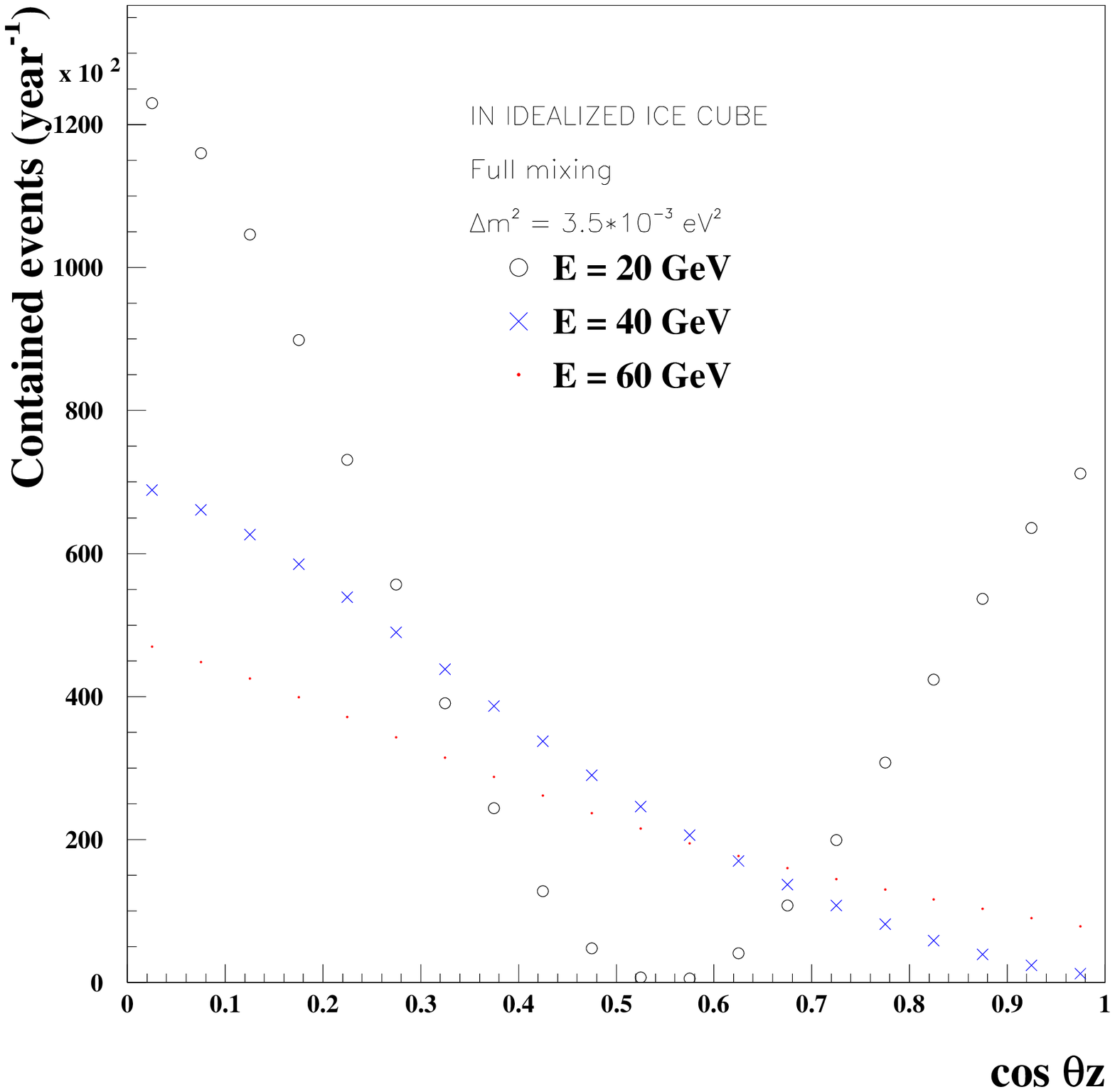}
\epsfxsize=225pt \epsfbox{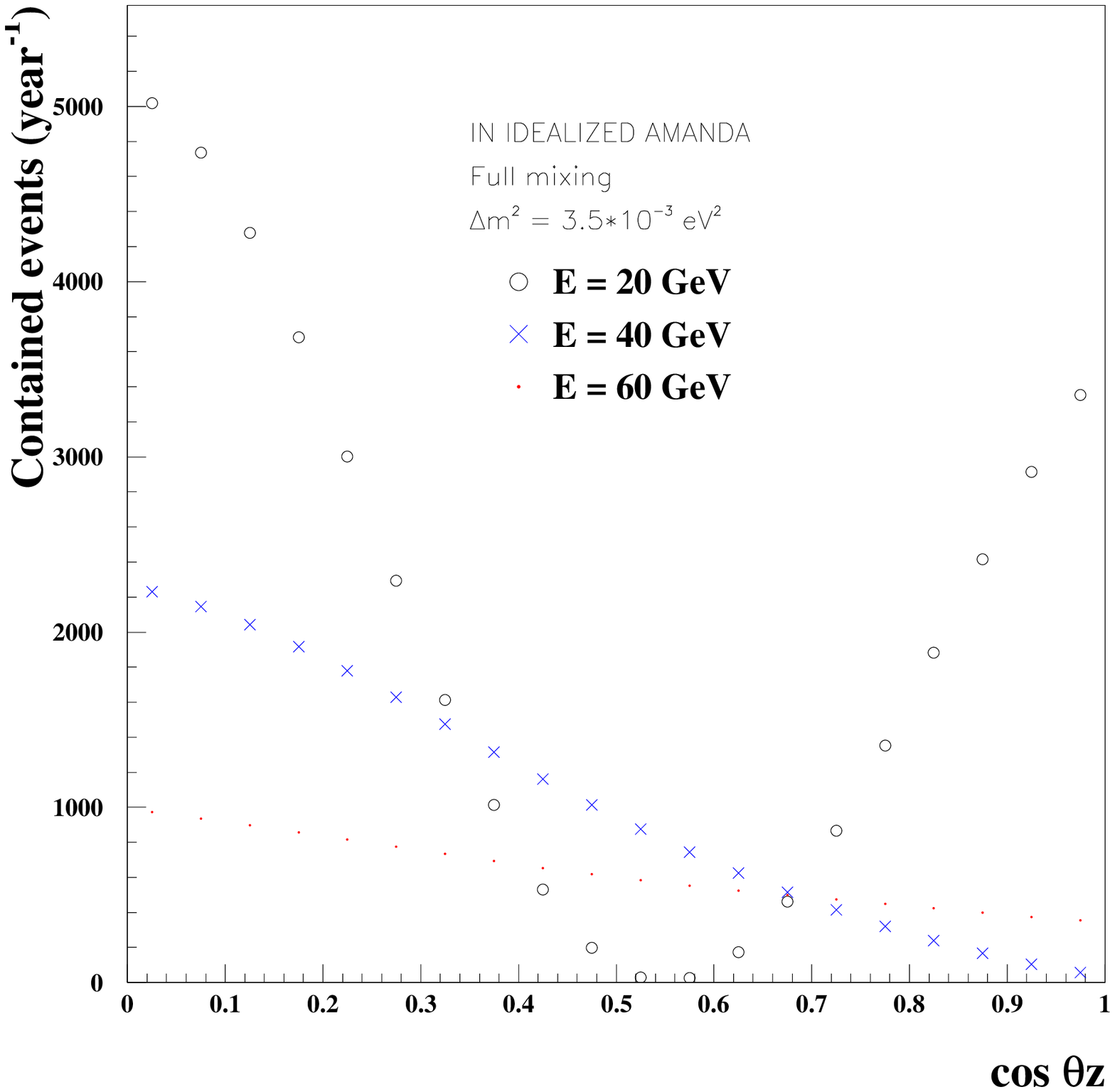} \\
\caption{Annual contained events in idealized IceCube and AMANDA (as labelled) in 
0.05 $\cos\theta_Z$ bins
versus cosine of the zenith angle. The neutrino energy is fixed as labelled.}
\label{fig:bin}
\end{figure}

Figure~\ref{fig:bin} shows the expected number of contained events per year 
for an idealized detector of the volume proposed for IceCube \cite{ice3} 
with height and diameter of a kilometer. 
The neutrino energy is fixed at 20, 40 or 60 GeV and $\cos\theta_Z$ divided in bins 
with a width of 0.05. Each bin corresponds to
about 0.3 sr. The same is shown  for the idealized detector with the size of AMANDA-II 
200-m diameter and 1-km height.

As up to now we are assuming an ideal detector, these figures show that the volume
of both AMANDA II and IceCube detectors are quite sufficient to measure atmospheric
neutrino oscillations. The key parameter for an idealized detector is that the energy
threshold has to be less than 30 GeV. The oscillation pattern for energies below this
threshold can be seen when measuring the number of contained events versus the neutrino
arrival direction.

However, instrumental effects have to be taken into consideration.
Before showing the sensitivity region for the $\Delta m^2$ and $\sin^2 2 \theta$
parameters that can be acheived with both AMANDA II and IceCube detectors, we will 
include these uncertainties in our analysis.

\section{Instrumental Effects and Limitations}
In this section we consider the degradation of the measurement that results 
from instrumental effects and limitations.
Two major issues are (1) angular and (2) energy reconstruction and resolution, 
since the physics manifests itself as a function of $L/E$ and we have $L$ 
as a function of $\cos \theta_Z$.

There are two kinds of analysis that can be done. One measures the number
of contained events with fixed energies (which is easy to visualize by measuring
the number of events versus $\cos \theta_Z$). The other measures the number 
of events versus the full $L/E$ spectrum.

To detect an oscillation pattern as a function of $L/E$ 
requires collecting enough events and determining $L/E$ to sufficient accuracy. 
From observations of the resulting muons and hadronic shower one estimates the 
incident neutrino arrival angle $\theta_Z$ and energy $E$.
The distance $L$ is determined from the angle $\theta$ (see Equation~\ref{eq:len}).
The fractional error in $L/E$ is given by:

\begin{equation}
\frac{\sigma_{L/E_\nu}^2}{(L/E)^2} = \frac{\sigma_{E_\nu}^2}{E_\nu^2} + 
\frac{\sigma_L^2}{L^2} 
\simeq \frac{\sigma^2_{E_\nu}}{E_\nu^2} + tan^2 \theta_Z \sigma_{\theta_Z^2}
\end{equation}
where $E_\nu$ is the neutrino energy
and in the approximation that $L = 2 R_\oplus \cos\theta_Z$
which is reasonably accurate for upward going events. 
For the next step we assume 
that the angle and energy correlation is small.
This is likely to be only partially correct but will show capabilities.

\subsection{Angular Reconstruction and Resolution}
\begin{figure}
\centering\leavevmode \epsfxsize=300pt \epsfbox{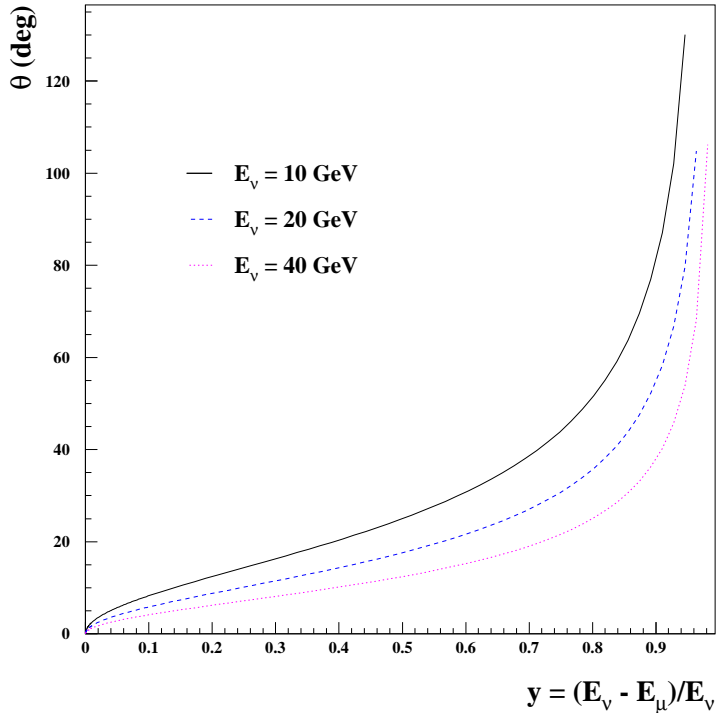}
\caption{Angle distribution for several energies (as labeled). $\theta$
corresponds to the angle between the neutrino and the muon directions.}
\label{fig:angdis}
\end{figure}

The muon does not travel and thus point in the same direction as 
the incident neutrino except when averaged over many events. 
This difference occurs because there is some exchange of energy (inelasticity) 
and momentum to the nucleon when the neutrino converts into a muon.
The inelasticity of the interaction and the spread in pointing
are related although not as simply as one would like.
Figure~\ref{fig:angdis} shows the distribution of the angle
$\theta$ between the muon and neutrino directions
for several relevant energies.
 
Characterized by one parameter, the RMS spread in direction 
between the incoming neutrino and 
the outgoing muon direction is given by the relation \cite{gaisser}:
\begin{equation}
\theta_{RMS} \simeq \sqrt{\frac{m_p}{E_\nu}}, 
\label{eq:deltheta}
\end{equation}
where the angle is given in radians.

\begin{figure}
\centering\leavevmode \epsfxsize=300pt \epsfbox{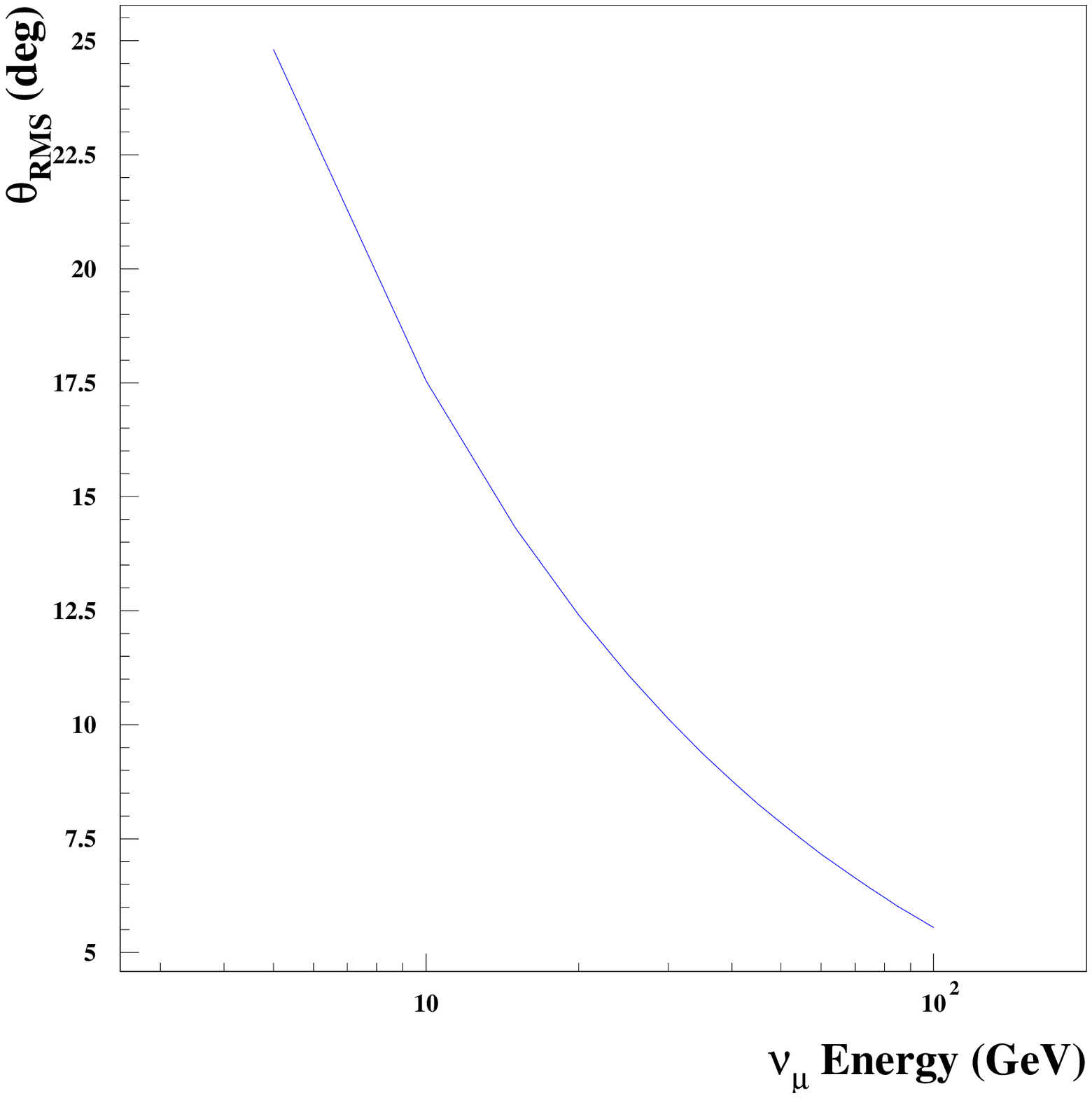}
\caption{Angle distribution as a function of neutrino energy. $\theta_{RMS}$
corresponds to the RMS angle between the neutrino and the muon directions.}
\label{fig:rms}
\end{figure}

The RMS for $\theta$ versus neutrino energy is shown in 
Figure~\ref{fig:rms}. This relation is valid for neutrino energies between
10 and 3000 GeV. Figure~\ref{fig:rms} shows that for neutrino energies of 
a few tens of GeV,
the angular RMS varies from about 8 to 17 degrees. The RMS is spread enough that
the direction of the neutrino will only be known by the average of the angular
distributions of many events. 

Figure~\ref{fig:avey} 
(extracted from \cite{quigg96}) shows the
mean inelasticity $<y>$ (where $y~=~(E_\nu~-~E_\mu)/E_\nu$) versus neutrino
energy. In a charged current interaction $<y>$ is almost constant between 10 and
100 GeV with a value of 0.48. Figure~\ref{fig:angdis} shows that this corresponds
to an average scattering angle of about 15 degrees for a 20 GeV neutrino.

The muon direction is determined by measuring the arrival time 
of the Cerenkov light at each phototube.
A muon of energy $E$ will have a mean range of $5E$~m/GeV. 
If timing is done to $\delta t$, the direction of the muon can 
be measured to about 
\begin{eqnarray}
\delta \theta = \frac{\sqrt{2} c \delta t }{5 E} \hspace*{.3cm} 
{\rm \frac{rad~GeV}{m}}  
= \frac{0.085 \delta t}{E} \hspace*{.3cm} {\rm \frac{rad~GeV}{ns}} = 
\nonumber \\
\frac{4.86^\circ \delta t}{E} \hspace*{.3cm} {\rm \frac{GeV}{ns}}
\end{eqnarray}
For approximately upward moving muons the direction can also be determined 
by timing arrival of photons along a vertical string of tubes and comparing it 
to the speed of light.
This technique measures $\cos \theta_Z$ directly and with reasonable parameters 
to an accuracy better than 0.1.

In the energy
range that we are interested in ($10 \leq E \leq 100 $ GeV) the angular resolution is
therefore dominated by the neutrino-muon scattering distributions.
In our analysis we will consider the average scattering angle. For more precision a
Monte Carlo analysis can be done, but we consider the average scattering angle precise
enough for the purpose of predicting the possibility of measuring neutrino oscillations.
Also, Figures~\ref{fig:surv}~and~\ref{fig:oscdm2} show that to observe a $\Delta m^2$ 
effect, observations do not have to be made near the horizon 
where the path length changes rapidly with angle. Thus
the intrinsic spread in path length due to angular error
is sufficiently small. 

One can therefore expect that for energies above about 10 GeV the angle 
of the incoming neutrinos can be determined 
sufficiently accurately for neutrino oscillation studies. 

\subsection{Energy Reconstruction and Resolution}
\label{sec:enres}

We now include uncertainties on the energy estimation of the incident neutrino. 
An ideal detector, that is equally and fully sensitive 
to all energy deposited in the detector, 
could reconstruct the incident neutrino energy very precisely 
for all fully contained events.

If all the energy from the neutrino went into the muon, 
then one could exploit the range energy relation and 
reconstruct the muon and thus the neutrino energy.

We can write the formula for the error in $E_\nu$ as
\begin{eqnarray}
\label{eq:sigmu}
\sigma_{E_\nu}^2 &=& \sigma^2_{E_\mu}
+ \sigma^2_{E_h} \nonumber \\
\frac{\sigma_{E_\nu^2}}{E_\nu^2} &=& \frac{\sigma^2_{E_\mu}}{E_\mu^2} 
(1 - y)^2 + 
\frac{\sigma^2_{E_h}}{E_h^2} y^2
\end{eqnarray}
where $E_\nu$, $E_\mu$, and $E_h$ are the neutrino, muon, and hadronic energies, 
and $y = E_h / E_\nu$ is the inelasticity of the interaction.
Note $E_\nu = E_\mu + E_h$, and $1 - y = E_\mu/E_\nu$.

The muon energy is determined by measuring its range - path length - 
which is on average 5 meters per GeV. 
Thus a determination of the path length to an accuracy of 5 meters 
corresponds to an error of 1 GeV independent of the energy of the muon.
The fluctuations in range have a slight energy dependence.

\begin{figure}
\centering\leavevmode \epsfxsize=300pt \epsfbox{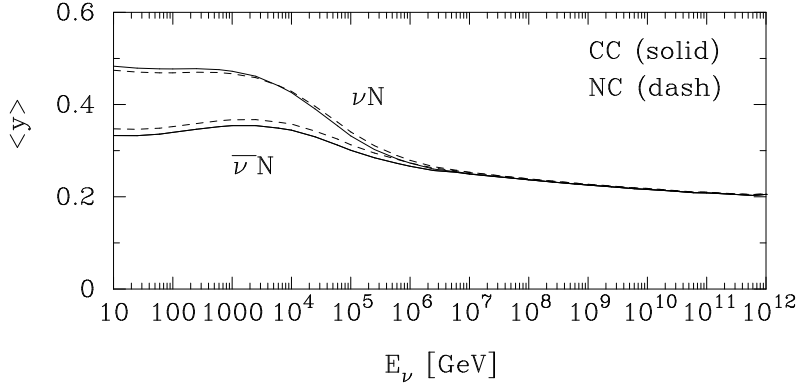}
\caption{Mean inelasticity versus energy. $<y> = (E_\nu - E_\mu)/E_\nu$.
Figure extracted from \cite{quigg96}.}
\label{fig:avey}
\end{figure}

One has to take the inelasticity in the conversion of neutrino to muons into
account.
Figure~\ref{fig:avey} (extracted from \cite{quigg96}) shows that
for energies between 10 and 1000 GeV the muon 
carries about 52\%\ of the incident neutrino energy. The inelasticity $y$
ranges nearly uniformly for the energy range of interest here. 
At much higher energies it decreases asymptotically to about 20\%\ inelasticity.

Thus to reconstruct the neutrino energy the detector must estimate both 
the muon energy (which is given by the muon range measurement and therefore
approximately independent of the energy of the muon) and the inelastic portion.
Roughly, the energy resolution is composed of a component which is nearly 
linearly dependent on energy and a component which is nearly independent
of the energy.
The muon range is estimated by the observations by optical detectors of the Cerenkov radiation as it travels through the ice (or water).
In general, the total light emitted is proportional to the energy deposited.
The muon is a line source of light while the other interaction products
are generally more localized and near the point of neutrino interaction.

The errors in calibration and fluctuations in the fraction of the energy 
transferred to the muon and to the other products of the interaction result 
in an error which is roughly proportional to the energy of the incident neutrino. 
This uncertainty actually decreases somewhat from very low energies to higher energies.

We can therefore rewrite Equation~\ref{eq:sigmu} as
the standard deviation of a function which depends on the neutrino energy: 
$\sigma_E^2 = x^2 E^2 + (\delta E)^2$ 
where $x$ is a constant between 0 and 1 and $\delta E \sim 1 GeV$, assuming
no correlation between these terms.  
The energy resolution is therefore a fraction of the energy value plus a constant. 
We give our results assuming 0, 10, and 20\% energy resolution.

\begin{figure}
\leavevmode \epsfxsize=225pt \epsfbox{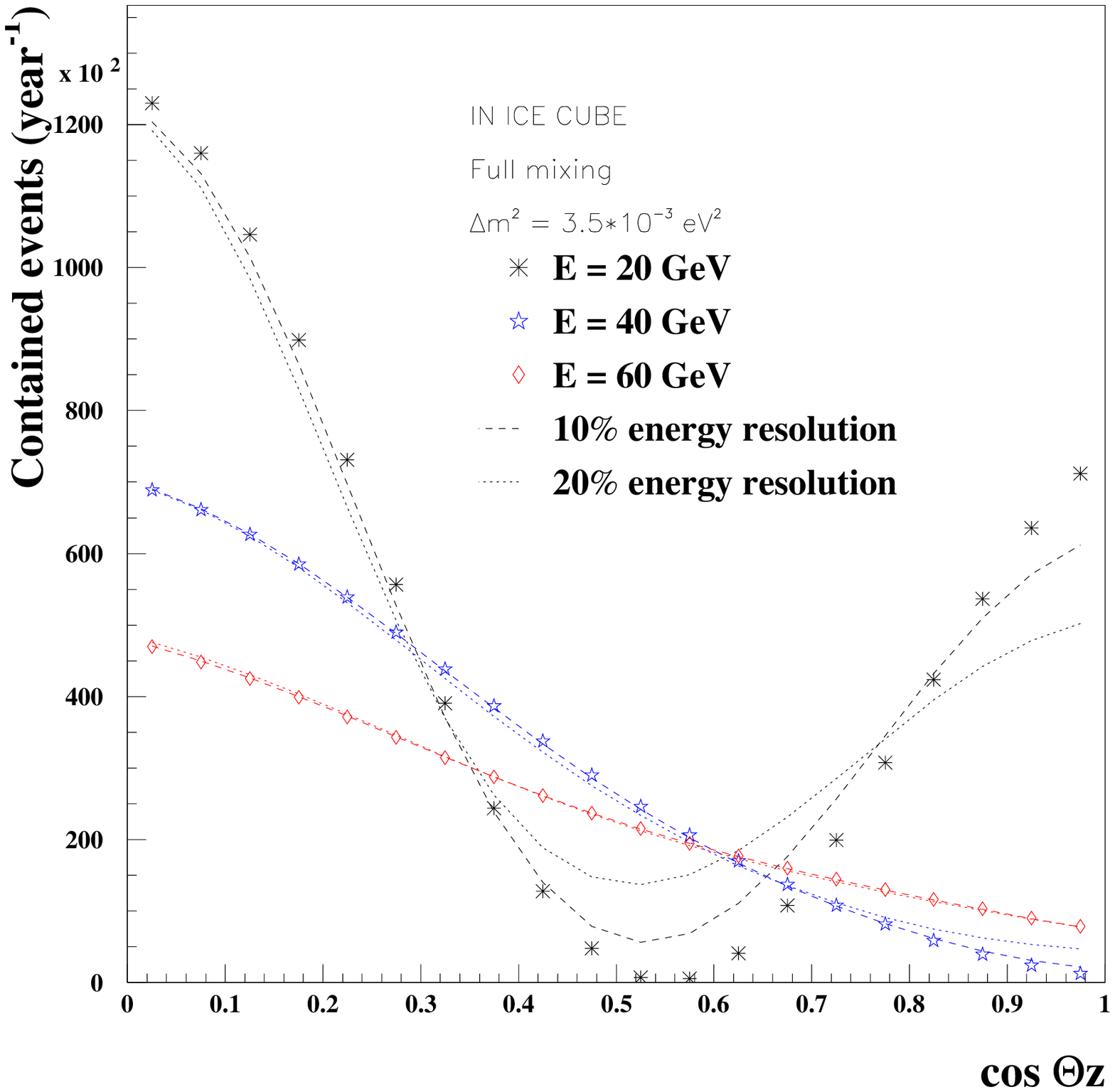}
\epsfxsize=225pt \epsfbox{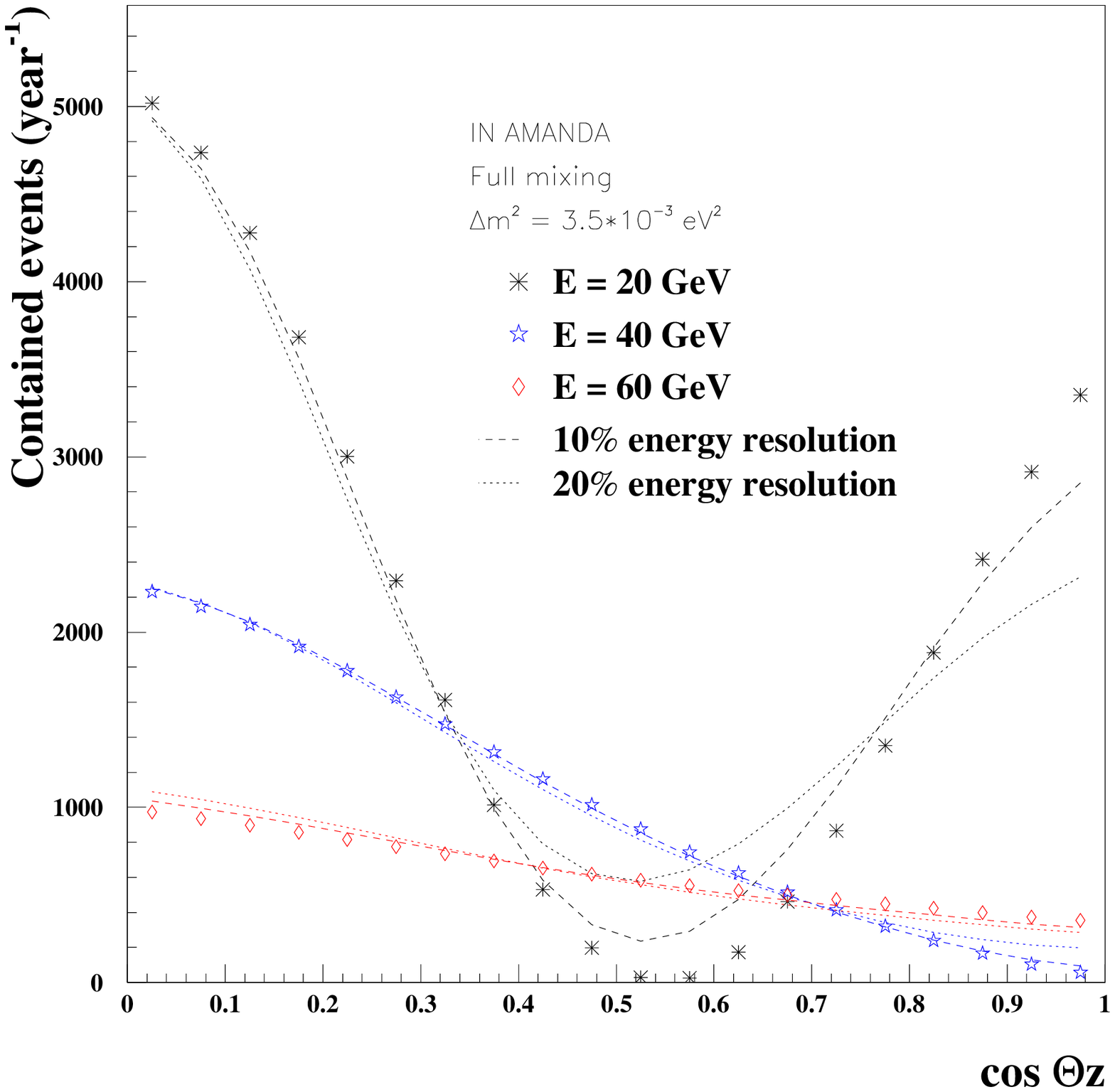} \\
\caption{Contained events in IceCube and AMANDA (as labelled) volume in 
0.05 $\cos\theta_Z$ bins. 
The neutrino energy is fixed as labelled and the effect of
10 and 20\% energy resolution is shown.}
\label{fig:bin_err}
\end{figure}

The number of contained events in IceCube and in AMANDA-II assuming the above energy
resolution are shown in Figure~\ref{fig:bin_err}. 
What one finds is that the standard deviation must be less than about 20\%\
of the energy (or FWHM less than about 50\%) in order to make quality oscillation 
observations.

The energy threshold is also important. To be able to observe the oscillation
pattern the energy threshold can not be much above 20 GeV. 

\begin{figure}
\centering\leavevmode 
\epsfig{figure=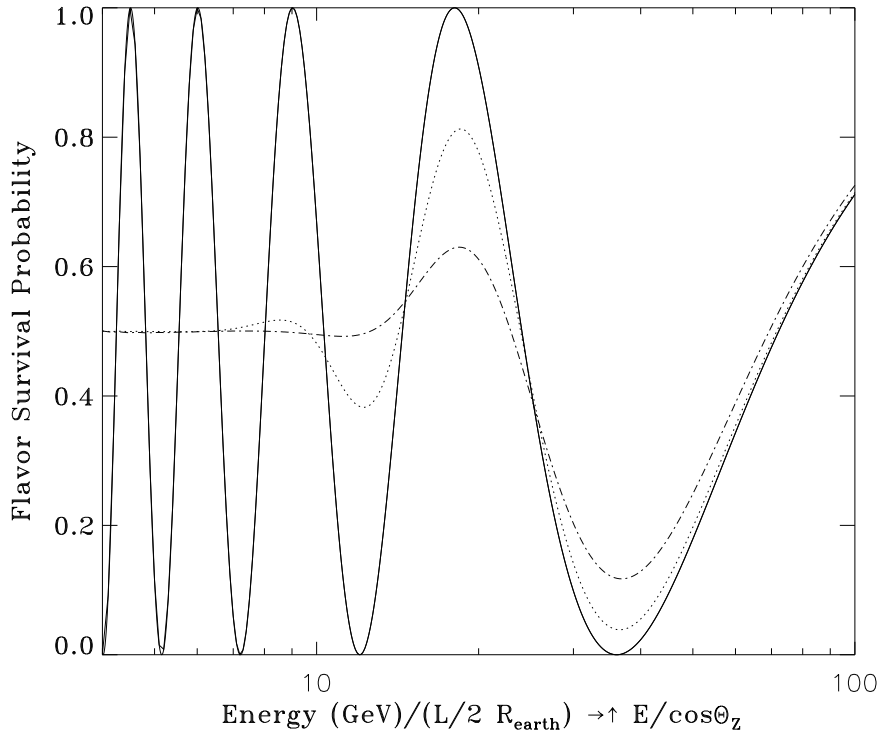,width=300pt,angle=90}
\caption{Flavor survival probability versus $E/\cos\theta_Z$.
Valid for upwards going neutrinos, where the distance traveled 
can be given in Earth radius units.
The dashed line corresponds to the effect of 10\% and the dotted
dashed to 20\% energy resolution.}
\label{fig:ecos}
\end{figure}

Figure~\ref{fig:ecos} shows the survival probability versus $E/L$ (or 
$E/\cos\theta_Z$) including 10 and 20\% energy resolution. 
In this kind of analysis,
vertical upwards events ($\cos \theta_Z = 1$) are the best for measuring neutrino 
oscillations.

IceCube \cite{ice3} is planned to have 81 strings in a $\rm km^3$ 
(125 m between each string). 
Each string will have 60 PMTs with 16 m between each optical module. 
AMANDA-II has 4 inner strings
with 20 PMTs in each of them; they are located in a cylindrical shape of
60 m diameter and 400 m length. There are also two outer set of PMTs; one
has 6 strings with 36 PMTs per string and 120m diameter (same height).
The outermost set has 9 more strings, 700 m long and instrumented every 14 m
(50 PMTs per string). The diameter of this outer string array is 200 m.

Both AMANDA and IceCube were designed to detect higher energies neutrinos.
In both experiments, for atmospheric neutrino oscillation measurements,
the strings and phototubes are too far apart to acheive the necessary energy
resolution. Also the detector energy threshold is set too high. This is true
for a fixed energy analysis (which would require a energy threshold around
20 GeV) and for the full energy spectrum (10-100 GeV) analysis.

For atmospheric neutrino oscillation measurements the current design proposed 
for IceCube has two major shortcomings:

\noindent
(1) Rejection of confusing events is too poor.
The individual strings of optical modules are too far apart 
to guard neighboring strings against muons coming near a string and appearing
to be a contained event or multiple muons depositing energy/light
near a string and simulating a contained event.
The rate of downward muons and muon bundles is about a million
times the contained event rate.

\noindent
(2) The vertical optical module spacing is also too great (by a small factor)
to provide sufficient energy resolution even for the vertical going 
contained events. 
One can estimate that with 16-m spacing that the range of the muon 
can be measured to about 8 to 10 meters RMS giving an error of about
2 GeV RMS for the muon energy. 
However about half the energy of the original neutrino energy is
deposited in the hadron shower but in a very wide distribution
which ranges basically from 0 to 100\%.
Thus it is necessary to determine how much energy is deposited inelastically,
which can be done by measuring how much light comes from the
hadronic/electronic shower to the optical module and then 
estimating how much energy was deposited by determing how far it was
from the shower to the optical modules.
With a single string the vertical distance of the main energy deposition
can be determined, though somewhat too poorly,  but the distance away from 
the string is much more poorly constrained.
For example, if the distance to the inelastic energy deposition
is determined to 2 to 3 meters out of a typical distance of 10 meters,
then the error in estimating the distance translates into an error
of 40\% or more in reconstructing the inelastic energy.
There is also an intrinsic spread in the amount of light received
due to the statistical fluctuations in transit to 16-m separated optical modules. 
The net result is that the total energy resolution for 
vertical going contained events is too poor by a significant amount
and is worse for other angles.

The current version of AMANDA-II also lacks the resolution 
necessary for atmospheric neutrino oscillation measurements.
At the present the trigger system and experimental procedures are 
set for measuring higher energy neutrinos with a high energy threshold 
(about 50 GeV) and discriminate against contained events.
 
AMANDA-II is closer to the necessary density of strings and optical modules
than IceCube, 
but the vertical spacings of 20 meters, 11 meters, and 14 meters and
average horizontal spacing of 40 meters is just too large to have
the required energy resolution. 
The set of six strings with 11-meter vertical spacing would be nearly 
adequate but need additional infrastructure in terms of other strings
nearby, calibration of the optical module response and the appropriate
triggering and data processing software.
Based upon the optical properties of the ice and optical modules
{\it in situ} and the performance of the detectors, we estimate that a 
spacing of order 5 to 10 meters would be adequate.

Similar arguments hold when it comes to utilizing AMANDA-II and IceCube
for future long baseline neutrino oscillation experiments \cite{Dicka}.
If the beams originate from CERN or Fermilab, latitudes about 46$^\circ$
and 43$^\circ$ north respectively, then a typical angle of the
incoming neutrino and produced muon is about 45$^\circ$ to the zenith
and there is a noticeable spread in the muon directions.
For that range of angles, the optical horizontal separation is as important
as the vertical separation. 
(Also note that the rough 45$^\circ$ zenith angle makes the effective
vertical separation $\sqrt{2}$ or 41\%\ greater.)
Though long-baseline experiments have the advantage of timing to reduce
backgrounds and thus enable lower energy thresholds set by the experimental
trigger \cite{Dicka},
the angles for likely beams lowers the energy resolution substantially.
The neutrino beam effective volume for IceCube would be quite low
due to the large spacing between strings.

\section{Tower Detector Configuration}
It is clear that measuring various values of $L$ and $E$ provides a very good 
and self-consistent test of neutrino oscillations.
However, it is possible to test and observe neutrino oscillations 
for a fixed distance, 
providing sufficient range of energy and number of neutrinos can be observed.

\begin{figure}
\centering\leavevmode 
\epsfig{figure=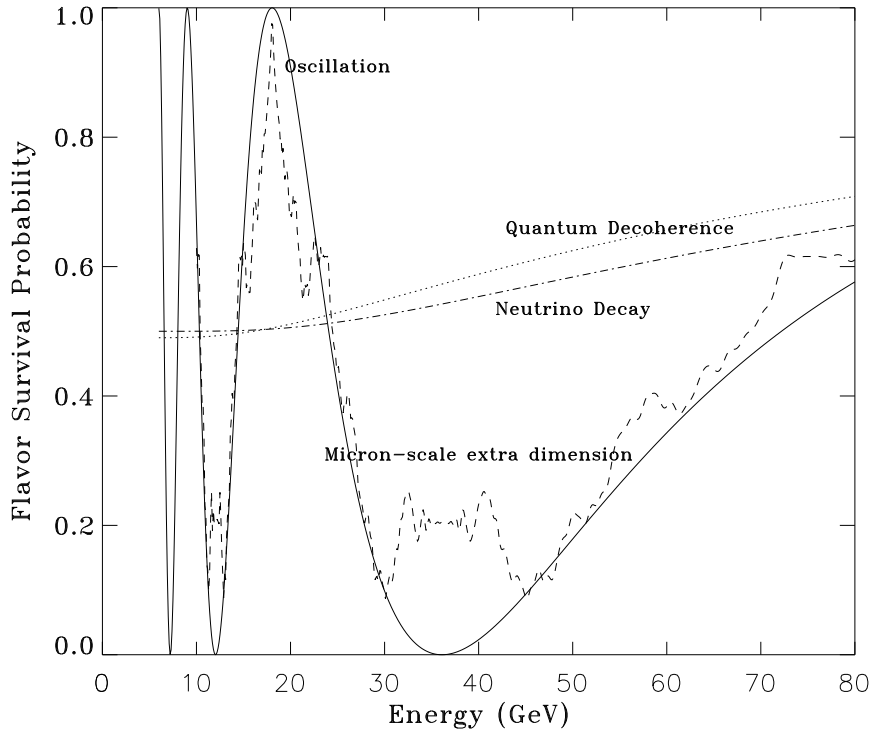,width=300pt,angle=90}
\caption{Muon neutrino survival probability versus neutrino energy
for upwards going neutrinos ($L = 2 R_\oplus$). Again four 
theoretical scenarios that might account for the observed effect are:
(1) Standard neutrino oscillations are shown by the solid line. 
Full neutrino mixing and \protect$\Delta m^2 = 3.5 \times 10^{-3}\; eV^2$ is 
assumed. 
(2) Dashed line is effect of additional dimension with characteristic radius 
of about a micron \cite{Barbieri}.
(3) Dashed Dotted line is effect of a decaying neutrino \cite{Barger}.
(4) Dashed line is the effect of quantum gravity decoherence \cite{Lisi}.}
\label{fig:moden}
\end{figure}

Figure~\ref{fig:moden} shows the muon survival probability versus
energy $E$ for upwards going neutrinos ($L = 2 R_\oplus$) superimposed by 
non-standard explanations.

We now consider explicitly a configuration in the shape of a long tower.
This tower is oriented vertically both to get the maximum path length 
$L \simeq 2 R_\oplus$ and 
because that is operationally the most reasonable configuration 
for water and ice-based detectors.
The tower geometry has its height much longer than its diameter so that 
its acceptance is near vertical (near direct upward and downward going directions) 
and its solid angle is quite limited. 
The limited solid angle means that the distance is effectively constant
(due to the slow change in $\cos \theta$ near 0 and 180 degrees).
Thus the blurring due to the change in $L$ adds 
very  little to the blurring caused by the energy resolution.

We consider the hypothetical configuration  
with optical detectors at 5-m to 10-m spacings along a string 1-km long.
The close optical detector spacing improves the energy resolution
both in terms of determining the muon range and in the energy deposited by 
the other interaction products.
The tower detector consists of four such strings
embedded in a larger detector, e.g. AMANDA-II.
The larger detector acts as an after the fact additional veto
and, when appropriate, provides additional information in constraining the event 
and its energy.

\begin{figure}
\centering\leavevmode 
\epsfig{figure=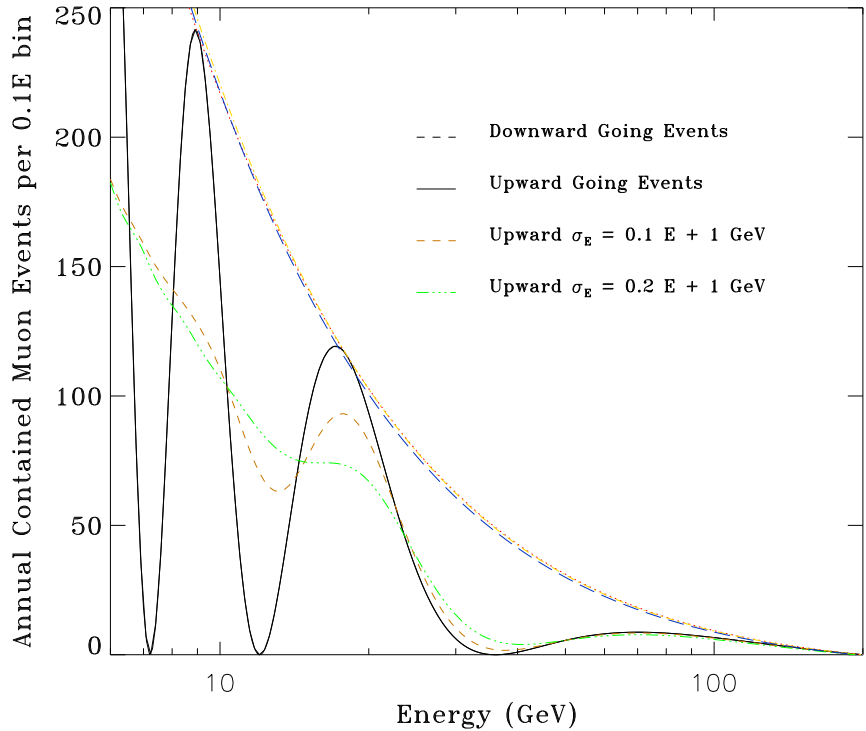,width=300pt,angle=90}
\caption{Annual contained events for tower configuration including 
a 10 and 20\% energy
resolution.}
\label{fig:enres}
\end{figure}

Figure~\ref{fig:enres} shows the number of contained 
events per year versus energy for vertical going neutrinos. 
In order to get the atmospheric muon neutrino flux at lower energies, 
we extrapolate the Volkova \cite{volkova} spectrum used 
in section~\ref{sec:nuflux} to 5 GeV. 
We compared this flux with the one calculated in \cite{agrawal} and 
they are in good agreement.



This figure shows the expected event rates as a function of energy 
with perfect energy resolution and including a 10 and
20\% energy resolution effect.

Previous figures in this section assume the measurement of the neutrino
energy through measuring the muon energy and hadronic energy. 
Another approach to observing neutrino oscillations is through the measurement 
of the muon range, without the information on the hadron energy
and reconstruction of the neutrino energy.
Such a measurement would have the advantage of a clean estimate
of muon energy.
Since the muon carries a variable fraction of the energy of its
parent neutrino, this is equivalent to having a random error in the
neutrino energy estimate with a bias to the low energy side.
This bias is not overwhelming because the atmospheric neutrino
energy spectrum is steeply falling and the number of muons
observed at energy $E_\mu$ is dominated by parent neutrino energies
which are just above the muon energies rather than those 
from much higher energies.

\begin{figure}
\centering\leavevmode 
\epsfig{figure=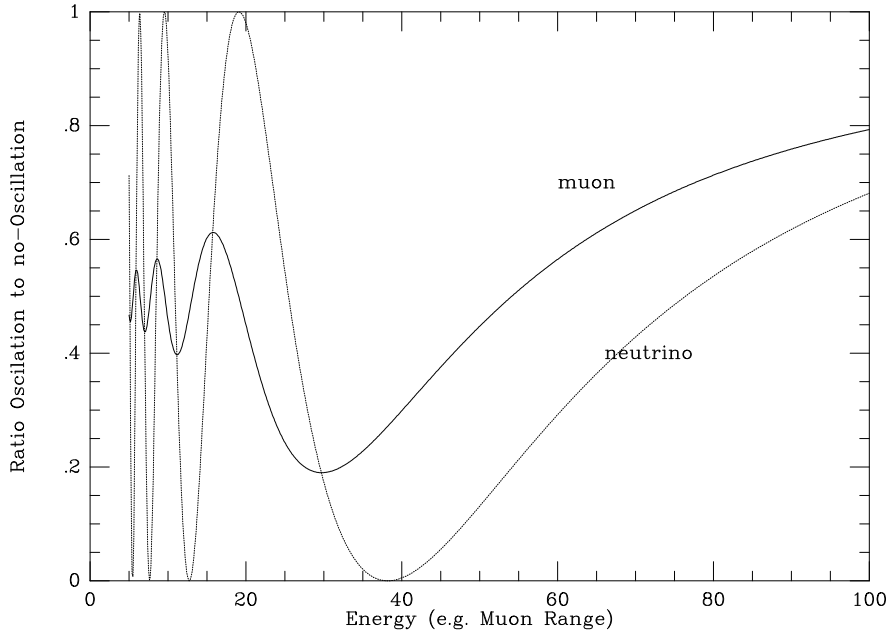,width=4in,angle=-90}
\caption{Ratio of the number of events
with oscillations over the number of events without oscillations 
versus the neutrino energy and versus the muon energy (GeV).}
\label{fig:muenr}
\end{figure}

Figure~\ref{fig:muenr} shows the ratio of the number of events
with oscillations over the number of events without oscillations 
versus the neutrino energy and versus the muon energy.
Measurement of neutrino oscillations using the muon energy information alone 
is still possible. 
This kind of analysis is proposed for ANTARES \cite{ANTARES} and the comparison
of Figure~\ref{fig:muenr} with Figure~\ref{fig:ant} shows that our
proposed tower configuration can measure neutrino oscillations as well
as ANTARES, if using only the muon energy information.
The key issue in that approach is to determine the ratio
of upward going versus downward going contained muon events.

\section{Backgrounds and Systematics}
This paper is not meant to be a complete analysis of the Tower configuration
for neutrino oscillation observations.
More work is necessary.
However, it is appropriate to outline and scope
anticipated backgrounds and systematic errors
to see both if there are any potential show stoppers and
which areas need further work.

\subsection{Atmospheric Muons}
The primary background is due to the very numerous downgoing muons
produced by cosmic rays hitting the atmosphere.
This is the same process that produces the atmospheric neutrinos.

A veto against these events puts a more stringent requirement than 
that needed to ensure that an event is contained.
A special trigger is required that looks for events and event topology
inside the active area of the detector and vetoes against particle
entering the detector volume from above and below or from the side.
An AMANDA style low multiplicity majority logic trigger
would be dominated by random coincidence from optical module noise.
Deadtime due to the veto is an issue so that separate triggers
would be necessary for separate science goals.
One must also be concerned about self-vetoing due to photons leaking 
from the real signal because the absorption length in ice is fairly long.
Overcoming this kind of self-vetoing requires careful timing 
which may in turn require after the fact processing of a much larger data set.

The observation must have very good rejection of muons coming in at an angle 
and thus appearing to be a contained event with very little inelasticity.
This is the primary motivation for locating the denser optical module strings 
inside the existing AMANDA-II array.
The combination of the Tower closed-packed and dense strings and 
the information from the 19 AMANDA-II strings provides
good rejection of well off-vertical muons.

\subsection{Electron Neutrinos}
At low energies (order of 10 GeV), electron neutrino charged-current interactions 
could mimic low energy muon neutrino induced events.
The fluxes are similar.  
Some are likely to be included in the data sample
and a careful study would be needed to show that they can
readily be rejected to the necessary level.
Fortunately, one can test the results using only the higher energy events
at some cost in sensitivity.

\subsection{Neutral Current Interactions}

Although Super-Kamiokande favors muon neutrino oscillation into tau
neutrino \cite{sktaufavor}, there is still the possibility that it
oscillates into sterile neutrinos. If the latter is the case,
muon neutrino neutral-current interactions are also a background to be 
considered. 
They will oscillate as the signal and due to an inelasticity comparable
to the charged current one (see Figure~\ref{fig:avey}) will produce
a jet of events from the nucleon recoil. This generates
some oscillation in the energy spectrum of this background.
The difficulty in distinguishing neutral-current and charge-current
events is that the muon range and the photon range in ice are similar.
So photons produced in the neutral current jet can reach the same optical 
modules as the muon Cerenkov light, 
and with fluctuations being what they are, it is easy to misreconstruct 
the vertex of the interaction and/or muon range.

Vertex reconstruction is very difficult for short range (50 m) muons
leaving the neutrino interaction vertex.
Without good vertex, the energy resolution degrades. 
To get energy resolution of order 20\%\ 
requires that the vertex is very well known.
This requires careful timing and calibration of the optical modules.

This background can be understood as long as the detection
and reconstruction efficiencies are the same as for the charged
current events.
Count rates will be lower with realistic reconstruction efficiencies, 
which at these energies is limited to some extent by the muon neutral
current interaction background.

\subsection{Tau-Neutrinos}
If there are neutrino oscillations,
as other explanations (as shown in Figure~\ref{fig:moden}), 
the muon neutrino, most likely they oscillate into tau neutrinos 
\cite{sktaufavor}. If this interpretation is correct, 
the tau neutrinos present a background which tends to wash out the 
oscillations.

The outgoing tau from the tau neutrino charged current interactions decays 
to muons about 18\%\ of the time. 
Thus the secondary muon will look like a muon neutrino charged current event
and attenuate the oscillation pattern.
This represents a slight decrease in sensitivity and is
comparable in effect to an energy degradation.

Another 11\%\ of taus decays into a pion plus a tau neutrino, 
and the pion may decay to a muon or occasionally mimic a muon.
Another 25\%\ of the taus decays in to a pion, pi-zero, and tau neutrino, 
and the pion decays into a muon.  
At 20 GeV, most pions interact before they decay.
To do a full evaluation of these effects requires
a simulation.
The same may be required to determine what fraction of these taus will
generate cascade backgrounds. 

The tau neutrino-interaction charged-current cross section has a threshold
around 3 GeV, so any tau background is quite energy dependent and may
cease to be important below 10 GeV.
This background could be quite noticeable degradation of the
signal for this experiment and for several of the others discussed here.


\section{Sensitivity to Parameters and Comparison with other Experiments}

\begin{figure}
\centering\leavevmode \epsfxsize=300pt \epsfbox{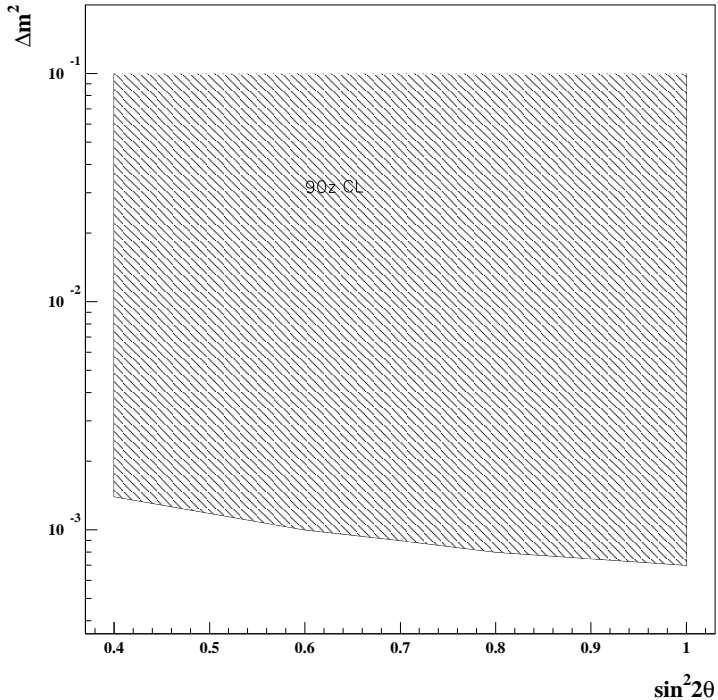}
\caption{$\sin^2 2 \theta$ versus $\Delta m^2$ sensitive region to demonstrate
neutrino oscillation with the proposed tower configuration. The shaded region
can be probed  with 
90\% CL and with a precision of 10\% or better in both $\sin^2 2 \theta$ 
and $\Delta m^2$. A detector energy threshold of 15 GeV and a 10\% energy
resolution is assumed.}
\label{fig:sens}
\end{figure}

Figure~\ref{fig:sens} shows the parameters 
$\sin^2 2 \theta$ and $\Delta m^2$ sensitivity region for the tower 
configuration described
above. This is the region where one can demonstrate oscillation with 
90\% CL and a precision of 10\% or better in both $\sin^2 2 \theta$ 
and $\Delta m^2$. We assume that the detector energy threshold is
15 GeV and the energy resolution is 10\%.
The sensitivity region for the tower configuration
is $\sin^2 2 \theta > 0.40$ and $\Delta m^2 > 1 \times 10^{-3}$.

\begin{figure}
\centering\leavevmode \epsfxsize=300pt \epsfbox{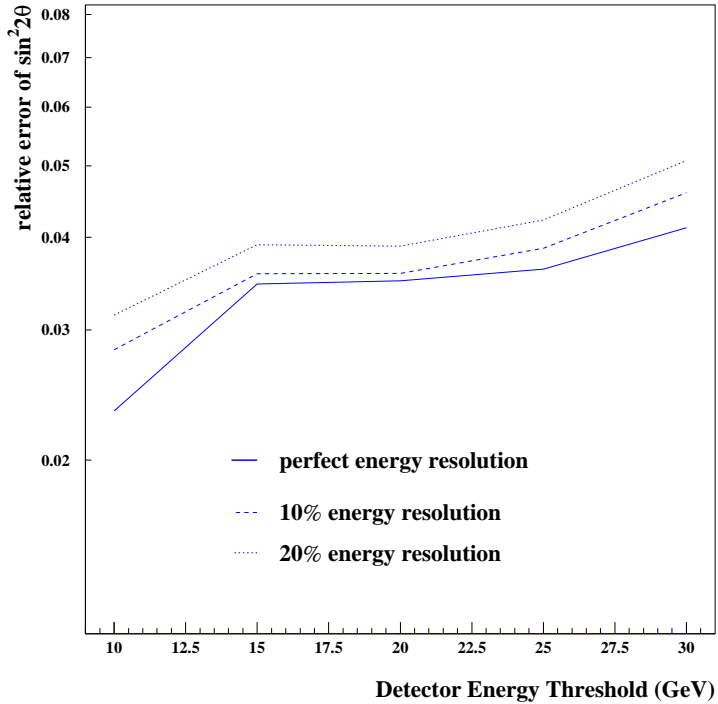}
\caption{90\% CL Estimated error in the mixing parameter $\sin^2 2 \theta$
versus detector energy threshold.This precision can be acheived around the
most probable value for neutrino oscillation ($\sin^2 2 \theta = 1$)
assuming the most probable value for $\Delta m^2$ ($\Delta m^2 = 
3.5 \times 10^{-3}$).}
\label{fig:sinerr}
\end{figure}

\begin{figure}
\centering\leavevmode \epsfxsize=300pt \epsfbox{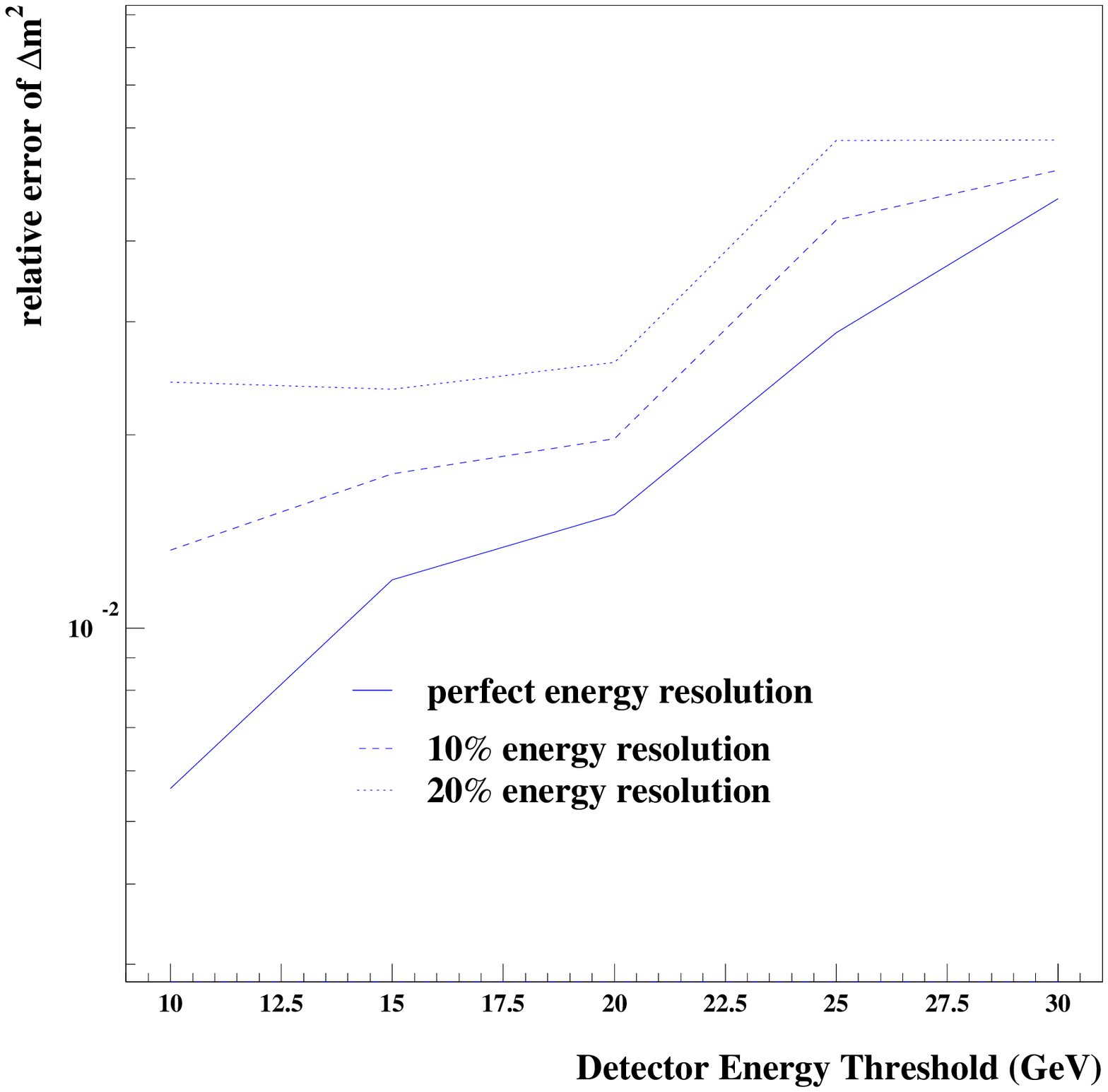}
\caption{90\% CL estimated error in the difference in the two neutrino mass 
squared $\Delta m^2$ versus detector energy threshold for the 
proposed tower configuration. This precision can be achieved around the
most probable value for neutrino oscillation ($\Delta m^2 = 
3.5 \times 10^{-3}$) assuming the most probale value for 
$\sin^2 2 \theta$ ($\sin^2 2 \theta = 1$).}
\label{fig:dm2err}
\end{figure}

A better precision can be achieved around the most
probable values of $\sin^2 2 \theta $ and
$\Delta m^2$ ($\sin^2 2 \theta = 1$ and 
$\Delta m^2 = 3.5 \times 10^{-3}$). This
is shown in Figures~\ref{fig:sinerr}~and~\ref{fig:dm2err}
for different detector energy threshold and energy resolution.
In this region the AMANDA II - Tower configuration expects
to have a relative error on $\sin^2 2\theta$ of order 3\% assuming the 
projected detector energy threshold of 15 GeV and 90\% CL. The relative error 
on $\Delta m^2$ is expected to be around 2\% for the same energy threshold
and confidence level.

The sensitivity region and the relative error of the parameters were 
determined by applying a maximum likelihood method \cite{pdg} assuming 
a Gaussian distributed data set and using Fisher matrix coefficients. 
This analysis does not include the effect of systematic errors nor 
correlates the parameters $\sin^2 2 \theta$ and  $\Delta m^2$. 
However, this is not a problem when comparing our proposed configuration with 
other detectors, 
since none of them include systematic bias in their sensitivity analysis.


A comparison with other experiments is shown in table~\ref{tab:exp}. 
MINOS \cite{minos} and MONOLITH \cite{monol} cover the broadest parameter region. 
All experiments including our proposed tower configuration have roughly the same 
sensitivity. 

An important point is that the experiments that compose Table~\ref{tab:exp} 
represent
three different techniques to measure neutrino oscillations. 
K2K \cite{k2ksen} and MINOS\cite{minos} use a controlled beam line and 
have two detectors, one close to the beam production
and another one far away. 
Although the detectors in each experiment are different from each other, 
the fact that they can control the beam and therefore have a good
energy and angular resolution is a strong characteristic of their design. 

\begin{table}
\caption[t1]{\label{tab:exp} Comparison among current or proposed neutrino 
experiments.}
\begin{tabular}{lccccccc}
\hline
Experiment & Energy & $\Delta m^2\ ({\rm eV^2})$ & $\sin^2 2\theta$ 
& Estimated   & Estimated & CL(\%) \\ 
& Threshold & & & Precision  & Precision & \\
& (GeV) & & & $(\Delta m^2)$ & $(\sin^2 2\theta)$ & \\
\hline
Proposed Tower & 16 & $> 10^{-3}$ & $ > 0.4$ & 10\% & 10\% & 90
        \\
K2K \cite{k2ksen} & 1 & $> 3 \times 10^{-3}$ &  $ > 0.4$ & 30\% & ? & 90
	 \\
MINOS \cite{minos} & 1 & $> 10^{-3}$ & $ > 0.1$ & 10\% & 10\% & 90
	 \\
MONOLITH \cite{monol} & 1.5 & $> 2 \times10^{-4}$ & $ > 0.2 $ & 6\% & ? & 90
          \\ 
ANTARES \cite{ant} & 5 & $> 10^{-3}$ & $> 0.6$ & 33\% & 33\%  & 99 \\
\hline
\end{tabular}
\end{table}

MONOLITH \cite{monol} is a massive and dense magnetized tracking calorimeter. The size and density of the detector are important to increase the number 
of neutrino interactions and the number of contained events. 
MONOLITH will cover a large range in $\Delta m^2$ since it is able 
to measure high momentum muons
which in other experiments is not possible 
since the muons range out of the detector before losing most of their energy.

The tower embedded in AMANDA-II that is proposed here and ANTARES are 
neutrino telescopes with large volumes in water or ice. 
Their advantage is their large size (detector volume). 
Both track and measure the energy of the muons from the Cerenkov light 
emitted in their passage through water or ice.

The current neutrino telescopes as AMANDA, Baikal and future telescopes as 
IceCube, NESTOR and NEMO have been aimed at detecting astrophysical neutrinos
and not designed specifically for neutrino oscillations.

ANTARES, in contrast, has been designed to be sensitive to neutrino oscillation 
measurements. ANTARES \cite{antares} has a smaller net volume than IceCube but 
has the potential 
to measure neutrino oscillations in the range of parameters indicated by 
Super-Kamionkande.

The fact that the neutrino oscillation hypothesis will be tested with different techniques
only enhances the importance of these experiments. 
All experiments will have a good sensitivity around the most probable values 
found by SuperKamiokande \cite{superk}.

Two other experimental possibilities are UNO \cite{uno} and 
a long baseline experiment
that would have a beam coming from either Fermilab or CERN CNGS to IceCube 
\cite{Dicka}.
UNO would be a Cerenkov detector with 20 times the volume of Superkamiokande.
It would identify electron and muon neutrino interactions
and would have good energy and vertex resolution as well 
as good reconstruction efficiency (comparable to SuperKamiokande).
UNO could do even longer baseline physics to look for CP violation.
However this would be in a much longer timescale and 
UNO would have to adjust its goals according to new results obtained.

The long baseline \cite{Dicka} that proposes a beamline from Fermilab or CERN
to IceCube would have to take into account the fact that the angle which the
beam would reach the detector is not favorable as described in section~\ref{sec:enres}.
Significant modifications would be necessary to the current design of IceCube 
to make the effort of sending a beam to IceCube work.

\begin{table}
\caption[t1]{\label{tab:time} Time scale for current or proposed neutrino 
experiments.}
\begin{tabular}{lcc}
\hline
Experiment & Start & Time scale to achieve \\
& Data Taking   & Expected Results \\ 
&    & (Years) \\
\hline
Proposed Tower & 2003 & 1 
        \\
K2K \cite{k2ksen} & June 1999 & 3
	 \\
MINOS \cite{minos} & 2003 & 2
	 \\
MONOLITH \cite{monol} & 2005-2006 & 4
          \\ 
ANTARES \cite{ant} & 2003 & 3 \\
\hline
\end{tabular}
\end{table}

Table \ref{tab:time} shows the quoted time scale for each experiment. K2K is the
only one already in data taking mode. All others will take two more years to start
collecting data.

MINOS will be able to cover a broad $\sin^2 2 \theta$ and $\Delta m^2$ parameter
space in a reasonable amount of time. Figure 4 of \cite{minos}
shows their precision
in these parameters assuming $\sin^2 2 \theta = 0.7$ and 
$\Delta m^2 = 5 \times 10^{-3} \;eV^2$. In the same figure they show simulated 
results where, assuming the above values,
they can distinguish among non oscillations and oscillations. The energy region 
shown
in this figure ranges from zero to 20 GeV. The region which probes the oscillation
pattern ranges from zero to less than 10 GeV.


Figure~\ref{fig:enres} shows that the Tower configuration can probe the region
above 15 GeV and determine the existence or not of the oscillation pattern.
Figure~\ref{fig:moden} shows that the region which can probe the oscillation
hypothesis versus non-standard scenarios that can account for the neutrino deficit,
is above about 10 GeV. So although MINOS can probe very well the current parameters
of neutrino oscillation, the tower configuration will be able to distinguish among
different scenarios that can account for the neutrino deficit. Even in the case 
that
the oscillation hypothesis holds, the Tower configuration and MINOS will complement
each other results since they are probing different energy ranges in about the same
timescale.

\begin{figure}
\centering\leavevmode \epsfxsize=300pt \epsfbox{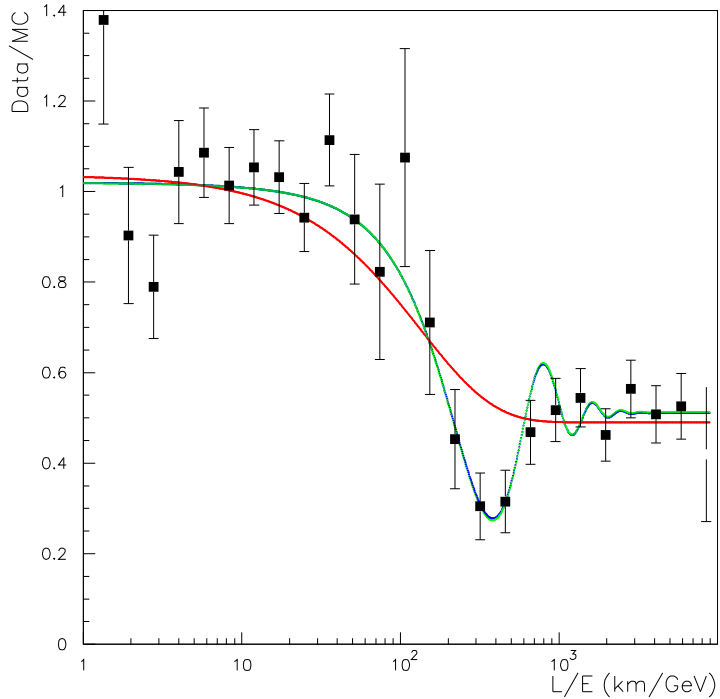}
\caption{L/E distribution expected from MONOLITH for $\Delta m^2 = 3 \times 10^{-3}\;
eV^2$ compared to the best fit oscillation hypothesis (oscillating line) and to
the best fit of the neutrino decay model. (extracted from MONOLITH proposal \cite{monol}.}
\label{fig:monol}
\end{figure}

Although MONOLITH can also probe a broad parameter space region, it can be 
considered
as another generation of neutrino experiments given their timescale for data 
taking.
Figure~\ref{fig:monol} (extracted from their proposal \cite{monol}) shows that they can 
also probe the 
existence of an oscillation pattern well above 10 GeV. In this way they will also
be able to distinguish among different models which can account for the SuperKamiokande
result. However their timescale is far beyond MINOS, K2K and the Tower 
configuration.

K2K also has good sensitivity for the oscillation parameter space and its
timescale is comparable with both MINOS and the tower configuration. 

\begin{figure}
\centering\leavevmode 
\epsfig{figure=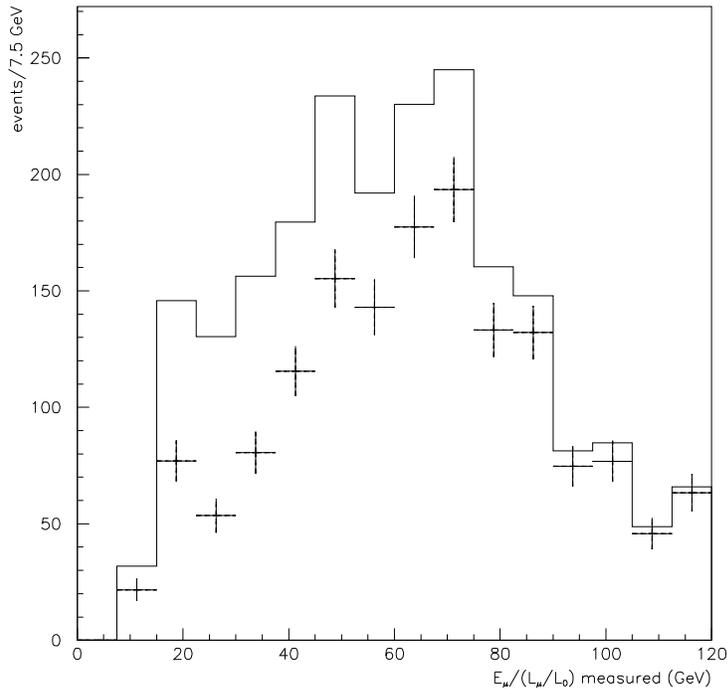,width=300pt}
\caption{ANTARES estimated number of reconstructed events versus $E/L$
from muon neutrinos undergoing oscillations with the parameters
$sin^2 \theta = 1$ (maximal coupling) and $\Delta m^2 = 0.0035$~eV$^2$
(points) and with no oscillations (histogram) for three years of observations.
(Extracted from Antares proposal \cite{ANTARES}).}
\label{fig:ant}
\end{figure}

ANTARES will also be able to test the oscillation hypothesis, 
since they will be able to go to higher energies.
Figure~\ref{fig:ant} shows the potential anticipated observations for 3 years by 
ANTARES \cite{ant}. 
Their time scale is longer than the Tower configuration
since they will need 3 years of data taking. 
Also the plot of their predicted results is based 
on the assumption that the atmospheric neutrino angular resolution can be measured to 
about 3 degrees.
We estimate that the intrinsic spread between the observed muon and 
incoming neutrino is of order 10 degrees.

The comparison made above shows that AMANDA-II modified to include the tower configuration
has a strong case for probing the neutrino oscillation hypothesis. 
Its sensitivity can assure a good constraint in the current allowed region 
for neutrino oscillations
and will achieve its goal in a short time scale when compared to other experiments.

One could argue that low energy physics should be done with lower
energy detectors and this is not the strong suit of IceCube or
AMANDA due to the photon scattering and low phototube density.
The argument here is that with some increase in phototube density
then the Cerenkov water detectors can be very effective
for atmospheric neutrino oscillation measurements because they can bring 
large volumes to bear.

\section{Summary}
High energy neutrino telescopes may be adapted to provide very powerful observations of
neutrino oscillations and/or some of the alternatives that might explain the up/down neutrino 
asymmetry observed by Super-Kamionkande and other anomalies.

\begin{acknowledgements}
This work supported by NSF Grants KDI 9872979 and Physics/Polar Programs 0071886
and in part by the Director, Office of
Energy Research, Office of High Energy and Nuclear Physics, Division of
High Energy Physics of the U.S. Department of Energy under Contract No.
DE-AC03-76SF00098 through the Lawrence Berkeley National Laboratory.
We thank Steve Barwick, Wick Haxton,Willi Chinowsky, John Jacobsen, Hitoshi Murayama 
and Howard Matis for comments.
\end{acknowledgements}


\begin{thebibliography}{99}
\bibitem{superk} Y. Fukuda {\em et al.}, Phys. Rev. Lett. {\bf 81}, 1562 (1998).

\bibitem{kamioka} K. S. Hirata {\em et al.}, Phys. Lett. {\bf B205}, 416 (1988); \\
  K. S. Hirata {\em et al.}, Phys. Lett. {\bf B280}, 146 (1992); \\ 
  Y. Fukuda {\em et al.}, Phys. Lett. {\bf B335}, 237 (1994).

\bibitem{imb} D. Casper {\em et al.}, Phys. Rev. Lett. {\bf 66}, 2561 (1991); \\
  R. Becker-Szendy {\em et al.}, Phys. Rev. {\bf D46}, 3720 (1992).

\bibitem{soudan} W. M. Allison {\em et al.}, Phys Lett. {\bf B391},  491 (1997).

\bibitem{Bahcall} J. N. Bahcall, P. I. Krastev, and A. Yu Smirnov, Phys. Rev. 
{\bf D 59},  046002 (1999).

\bibitem{Fogli} G. L. Foglie {\em et al.}, Phys Rev. Lett. {\bf 82},  2640 (1999).

\bibitem{Barger} V. Barger {\em et al.}, Phys Rev. {\bf D 57},  5893 (1998).

\bibitem{Barbieri} R. Barbieri, P. Creminelli, and A. Strumia, hep-ph/0002199 (2000).

\bibitem{Lisi} E. Lisi, A. Marrone, and D. Montanino, hep-ph/0002053 (2000).
 
\bibitem{frejus1} Ch. Berger {\em et al.}, Phys. Lett. {\bf B245}, 305 (1990).

\bibitem{nusex} M. Aglietta {\em et al.}, Europhys. Lett. {\bf 8}, 611 (1989).

\bibitem{kayser} P. Fisher, B. Kayser, K. S. Macfarland, Ann. Rev. Nucl. Part. Sci.
{\bf 49}, 481 (1999).

\bibitem{sktaufavor} S. Fukuda {\em et al.}, Phys. Rev. Lett. {\bf 85} 3999 (2000).

\bibitem{amanda} E. Andres {\em et al.}, Astropart. Phys. {\bf 13}, 1 (2000).

\bibitem{ice3} See the proposal for ICE CUBE at http://pheno.physics.wisc.edu/icecube/ .

\bibitem{antares} astro-ph/9907432 (1999).

\bibitem{nestor} Nucl.Phys.Proc.Suppl.{\bf 70},442 (1999).

\bibitem{volkova} L. V. Volkova, Yad. Fiz. {\bf 31}, 1510 (1980). Also published at
Sov. J. Nucl. Phys. {\bf 31}, 784 (1980).

\bibitem{honda} M. Honda, T. Kajita, K. Kasahara and S. Midorikawa, 
Phys. Rev. D {\bf 52}, 4985 (1995).

\bibitem{agrawal} V. Agrawal, T. K. Gaisser, P. Lipari,P. and T. Stanev, 
Phys. Rev. D {\bf 53}, 1314 (1996).

\bibitem{quigg} R. Gandhi, C. Quigg, M. H. Reno and I. Sarcevic, 
Phys. Rev. D {\bf 58}, 093009 (1998).

\bibitem{lip} P. Lipari, M. Lusignoli and F. Sartogo, Phys. Rev. Lett. {\bf 74}, 4384 (1995).

\bibitem{pdg} Particle Data Group, The European Phys. Jour. {\bf 15}, section 23.6
(2000).

\bibitem{quigg96} R. Gandhi, C. Quigg, M. H. Reno and I. Sarcevic, Astropart. Phys. {\bf 5},
81 (1996).

\bibitem{gaisser} T. Gaisser, ``Cosmic Rays and Particle Physics'', Cambridge University
Press (1990).

\bibitem{ANTARES} ANTARES web page, http://antares.in2p3.fr/, see, for example, the ANTARES proposal and related publications.

\bibitem{Dicka} K. Dick, M. Freund, P. Huber, and M. Linder, hep-ph/0008016 (2000).

\bibitem{ant} C. Carloganu, Europhysics Neutrino Oscillation Workshop (NOW 2000), 
Otranto, Italy (2000).

\bibitem{k2ksen} Y. Oyama, Talk given at the YITP Workshop on Flavor Physics, Kyoto, Japan (1998),
hep-ex/980314 (1998). //
S. Boyd, Talk given at the Sixth International Workshop on Tau Lepton Physics, 
Victoria, Canada  (2000), hep-ex/0011039.

\bibitem{minos} D. A. Petyt, Phys. of Atomic Nuclei {\bf 63}, 1122 (2000).

\bibitem{monol}P. Antonioli, Europhysics Neutrino Oscillation Workshop (NOW2000), Otranto, Italy
(2000), hep-ex/0101040.\\
MONOLITH Proposal - CERN/SPSC 2000-031.

\bibitem{uno} C. K. Jung, astro-ph/0005046 (2000).

\end{thebibliography}
\end{document}